\documentstyle[prb,aps,epsf,eqsecnum,multicol]{revtex}
\begin{document}

\draft
%\tighten

\title{Nonlinear Hydrodynamics of {\it Disentangled} Flux-Line Liquids}

\author{Panayotis Benetatos and M. Cristina Marchetti}
\address{Physics Department, Syracuse University, Syracuse, NY 13244}

\date{\today}

\maketitle

\begin{abstract}
In this paper we use non-Gaussian hydrodynamics to study the magnetic response
of a flux-line liquid in the mixed state of a type-II superconductor.
Both the derivation of our model, which goes beyond conventional Gaussian 
flux liquid hydrodynamics, and its relationship to other approaches used in 
the literature are discussed.
We focus on the response to a transverse tilting field
which is controlled by  the tilt modulus, $c_{44}$, of the flux array.
We show that interaction effects can enhance $c_{44}$ even in infinitely
thick clean materials. This enhancement can be interpreted as the
appearance of a disentangled flux-liquid fraction.
In contrast to earlier work, our theory
incorporates the nonlocality of the intervortex interaction in the field direction.
This nonlocality  is crucial for obtaining a nonvanishing renormalization
of the tilt modulus
in the thermodynamic limit of thick samples.

\end{abstract}

\begin{multicols}{2}
%\narrowtext
%\widetext
\section{Introduction}

The static and dynamical properties of magnetic flux lattices in type-II
superconductors have been the focus of much theoretical and experimental
work over the last ten years \cite{blatter,crab}. Interest in this field was revived by
the discovery of the high-$T_c$ materials, where thermal fluctuations melt the Abrikosov flux
lattice at temperatures and fields well below the mean field transition at
$H_{c2}(T)$ \cite{nelson,seung}.  The flux lattice melting is a first order
transition in clean samples\cite{Brezin}, with an associated jump in the bulk
magnetization, and it has been observed experimentally 
\cite{meltexI,meltexII,meltexIII,meltexIV,meltexV,meltexVI,meltexVII,meltexVIII}. 
In conventional low-temperature type-II superconductors, the region of the
phase diagram where thermal fluctuations are important is extremely small
and mean field theory provides a good description of the physics of the
flux-line array.  In the high-$T_c$ materials, in contrast, the melted flux
liquid replaces the Abrikosov lattice over a large region of the phase
diagram. Understanding the properties of the flux liquid is therefore
crucial for controlling the magnetic response of these materials.

The conventional Abrikosov flux lattice is characterized by two broken
symmetries.  First, the translational symmetry is broken by the ordering of
the magnetic flux lines in a triangular lattice in the plane perpendicular
to the external field.  Secondly, the gauge symmetry along the field is
broken by the alignment of the vortices with the external field. A natural
question then arises of whether these two symmetries are recovered
simultaneously upon melting, or rather they are recovered in succession at
two different temperatures. The latter scenario would allow for the
appearance of a disentangled flux liquid phase where translational symmetry
is recovered, but the longitudinal gauge symmetry is still broken. At a
second transition temperature the disentangled flux liquid would then be
replaced by an entangled flux liquid where the longitudinal gauge symmetry
is also recovered. Alternatively, if both symmetries are recovered
simultaneously, the Abrikosov lattice would melt directly into an entangled
flux liquid. The precise nature of such an entangled liquid 
remains an open question \cite{foota}.  The
existence of a disentangled liquid phase, exhibiting longitudinal
superconductivity {--} the ability to support currents flowing without
dissipation in the direction parallel to the flux lines {--} in clean samples
has been proposed some time ago by Feigel'man and collaborators
\cite{feigel}. Early simulations provided support for Feigel'man's ideas
\cite{liteitel,chenteitelI,chenteitelII}, but more recent numerical work indicates that the two
transitions observed in earlier work may have been the consequence of
finite size effects \cite{Nguyensudbo,japan}. Recent numerical results support
the scenario that the Abrikosov lattice melts directly into an
entangled liquid and no disentangled liquid phase exists in infinitely
thick samples \cite{Nguyensudbo,japan,nordborg}.  Open questions, however, remain concerning the role of
various approximations used in the different numerical models, 
particularly the range of the intervortex interaction.

A closely related property of the vortex array that provides a direct 
measure of longitudinal vortex correlations is the tilt modulus,
$c_{44}$. It can be probed  by
measuring the response of the flux array to a small additional magnetic
field $\delta{\bf H}_\perp$, applied perpendicular to the external 
field $\hat{\bf z}H_0$ responsible for the onset of the vortex state. 
Such a transverse field tilts the lines away from the direction of 
alignment with $H_0$. Correlated disorder induced, for instance, by aligned damage tracks
in the material can drive $1/c_{44}$ to zero, yielding a 
transverse Meissner effect, which has
been proposed as the signature of the Bose glass phase \cite{drnvv,trmeissexp}.
The role of correlated disorder in enhancing $c_{44}$ in the
liquid phase has also been observed
experimentally in materials with a single family of twin planes by using the dc flux transformer
configuration\cite{lopezI}. These materials contain practically no small-scale
disorder, so that the macroscopic flux liquid regions in the channels between twin planes
are very clean. The experiments suggest that the enhancement of $c_{44}$, interpreted as the
onset of disentangled liquid phase, be 
a finite-size effect, that decreases with increasing sample thickness \cite{lopezII}. 
In thick samples the experiments indicate that 
the vortex-lattice melting and the loss of longitudinal superconductivity coincide
in clean materials. Even though a true Meissner effect with vanishing $1/c_{44}$ is not
expected in infinitely thick, clean samples,
it is clear that 
interactions can enhance the tilt modulus of clean flux liquids and suppress the transverse 
response of the superconductor. 

In this paper we employ hydrodynamics to evaluate the renormalization of the tilt modulus
of a clean flux liquid due to interactions. Our starting point is a long-wavelength 
hydrodynamic free 
energy that includes {\it non-Gaussian} couplings in the hydrodynamic fields. 
It therefore goes beyond the Gaussian flux-line liquid hydrodynamic free
energy discussed before in the literature \cite{mcmhyd,mcmphysica}. 
We show that such a non-Gaussian hydrodynamic free energy can either be 
written down phenomenologically or it can be derived by using the mapping
of the classical  statistical mechanics of vortex lines with  {\it nonlocal} interactions
onto the quantum statistical mechanics of two-dimensional {\it charged} bosons,
introduced some time ago by
Feigel'man and collaborators\cite{feigel}.
Our central result is the expression for the renormalized {\it wave vector-dependent}
tilt modulus given in Eq. 1.7 below. This is a perturbative result 
that extends earlier results by other 
authors \cite{TN,LV} in two important ways. 
First, it incorporates both
the finite range and the nonlocality of the intervortex 
interaction in the field direction. This nonlocality plays a crucial role in controlling
the tilt response. It is only when the nonlocality is properly accounted for that
a finite renormalization of $c_{44}$ is obtained
in a clean flux-line liquids of infinite thickness.
In addition, our formalism allows us to evaluate the full 
wave vector dependence 
of the renormalized tilt modulus - a result that  was not discussed  before in the literature.

Before discussing our result in more detail, it is useful to make contact 
with already existing work.
The tilt modulus of  the Abrikosov lattice is easily calculated  from the Ginzburg-Landau 
free energy for a superconductor in a field. It is dispersive both in the
longitudinal and in the in-plane directions due to the non-local character
of the intervortex interaction and it has a rather complicated expression,
particularly for layered material. It naturally separates in the sum
of two contributions,
\begin{equation}
\label{tilt}
c_{44}(q_\perp, q_z)=c_{44}^v(q_z)+c_{44}^c(q_\perp, q_z),
\end{equation}
with $q_\perp$ and $q_z$ wave vectors perpendicular and parallel to the 
external field, respectively. The first term on the right hand side of Eq. (\ref{tilt})
is the single vortex contribution, arising from the self-energy part of the tilt energy.
Neglecting its weak logarithmic dependence on $q_z$, it is given 
by \cite{brandt,brandtI,brandtsudbo,fishertilt}
\begin{equation}
\label{tiltv}
c_{44}^v\approx n_0\tilde{\epsilon}_1,
\end{equation}
where $n_0=B_{0z}/\phi_0$ is the average areal density of vortices, with $B_{0z}$ the mean induction along
the external field direction and $\phi_0=hc/2e$ the flux quantum, and 
$\tilde{\epsilon}_1$ is the single-vortex tilt energy defined below.
The second term in Eq. (\ref{tilt}) is the compressional contribution from intervortex interactions.
It is strongly dispersive and in layered materials it is given 
by \cite{brandtI,brandtsudbo,fishertilt}
\begin{equation}
\label{tiltc}
c_{44}^c(q_\perp,q_z)= {B_{0z}^2\over 4\pi} \frac{1}{1+q_z^2
 \tilde{\lambda}_{\perp}^2 + q_{\perp}^2 p^2
\tilde{\lambda}_{\perp}^2},
\end{equation}
where $\ \tilde  \lambda_{\perp} = \lambda_{\perp} / (1-H/H_{c2})^{1/2}\ $ 
is the effective penetration length in the  $\ ab\ $ plane (the field is applied along the
$\ \hat{c}\ $ axis) and $p$ is the anisotropy ratio.
It is important to stress that the long wavelength tilt modulus,
\begin{equation}
\label{tilt0}
c_{44}=c_{44}(q_\perp=0,q_z=0)={B_{0z}^2\over 4\pi}\Big(1+{1\over 4\pi\tilde{\lambda}_\perp^2p^2n_0}\Big)
\end{equation}
is generally dominated by the large compressional contribution ($B_{0z}^2/4\pi$).
The second term inside the brackets in Eq. (\ref{tilt0}), 
arising from the single-vortex contribution,
is important only at very low vortex densities.

The tilt modulus of a {\it flux-line liquid} cannot be evaluated directly.
It is, however, expected that the bare flux-liquid tilt modulus, denoted here by
$c^0_{44}(q_\perp,q_z)$, does not differ considerably
from that of the lattice given in Eq. (\ref{tilt}) \cite{mcmdis}.
In fact, a direct coarse-graining of the microscopic
intervortex interaction yields a Gaussian long-wavelength free
energy of an entangled flux-line liquid with a tilt modulus 
given precisely by Eq. (\ref{tilt}) above \cite{mcment}.
Interactions responsible for nonlinearities
in the long-wavelength free energy will, however,
renormalize $c_{44}^0$. 

The renormalization of $c_{44}$ in flux-line liquids
has been studied before by employing the
analogy between the directed vortex lines induced in a three
dimensional superconductor by the external field $\hat{\bf z}H_0$ 
and the imaginary-time world lines of two 
dimensional bosons \cite{fisher,nelson,seung}. 
The most severe approximation made in the form of this  boson mapping
introduced by Nelson \cite{nelson,seung}, is that the pairwise interaction between flux lines
is approximated as local in the field direction ($z$), i.e., only the interaction between 
vortex segments at equal height $z$ is considered. This corresponds to an 
instantaneous pairwise interaction between the bosons. 
One of the consequences of this approximation is that it completely
neglects the compressional part of the tilt modulus. Hence in this model
$c_{44}$ is given by the single vortex part,
which is inversely
proportional to the boson superfluid density, $n_s$, 
\begin{equation}
\label{TNtilt}
c_{44}^v={B_{0z}^2\over 4\pi}{1\over 4\pi\lambda^2_\perp p^2n_s}.
\end{equation}
The superfluid phase of bosons ($n_s=n_0$) corresponds to an 
entangled liquid of magnetic flux lines with $c_{44}^v$  given by Eq. 
(\ref{tiltv}).
A finite normal-fluid fraction of bosons of density $n_n=n_0-n_s$ corresponds to a disentangled fraction of
flux liquid and enhances the tilt modulus. A normal-fluid phase of bosons
with $n_s=0$ corresponds to a disentangled flux liquid with infinite
tilt modulus and transverse Meissner effect. 
T\"auber and Nelson (TN) recently employed this boson mapping to evaluate the 
renormalization of $c_{44}^v$ due to 
sample thickness, different boundary conditions and various 
type of disorder \cite{TN}. They found that for finite sample thickness
(corresponding to a nonzero boson temperature) 
there is a nonvanishing normal-fluid component which suppresses $c_{44}^v$.
On the other hand, the normal-fluid density always vanishes for infinitely
thick samples (or vanishing boson temperature), 
so that the flux liquid is always entangled
in this limit.

Feigel'man and coworkers \cite{feigel} incorporated 
the nonlocality of the intervortex interaction in the field direction in the boson 
formalism. They showed that the statistical mechanics of vortex lines
with {\it nonlocal} interactions maps onto that of two-dimensional
{\it charged} bosons. This nonlocal mapping incorporates the compressional part of 
the vortex tilt modulus.
Larkin and Vinokur \cite{LV} and later Geshkenbein \cite{geshkenbein}
used this nonlocal boson
mapping to generalize the expression (\ref{TNtilt})
obtained by TN. These authors proposed that  the
long-wavelength renormalized tilt modulus can be written in terms of
the superfluid density $n_s$ of two-dimensional  bosons 
interacting with a gauge field as
\begin{equation}
\label{LVresult}
c_{44}^{LV}={B_{0z}^2\over 4\pi}\Big(1+{1\over 4\pi\tilde{\lambda}_\perp^2p^2 n_s}\Big).
\end{equation}
The superfluid density was evaluated perturbatively 
by Feigel'man and coworkers \cite{feigel}
for the case where the repulsive interaction among the bosons 
is infinitely long-ranged, corresponding to a vortex liquid with
$\lambda_\perp\rightarrow\infty$.   These authors argued that in this limit 
a distinct
disentangled flux liquid phase with diverging $c_{44}$  exists in 
infinitely thick superconducting samples.

The calculation of the interaction-renormalization of the flux liquid
tilt modulus via  hydrodynamics described here has
the advantage that it naturally incorporates the
nonlocality of the intervortex interaction and it allows us to easily treat
the case of finite $\lambda$. The non-Gaussian hydrodynamics used 
as the starting point
contains bare elastic constants that are determined by the intervortex interaction.
In particular, the bare tilt modulus is given by Eq. (\ref{tilt}). 
The corrections to $c_{44}$ due to the nonlinearities are evaluated perturbatively.
Our main result is an expression for the wave vector-dependent renormalized tilt
modulus,
given by
\begin{equation}
\label{result}
{1\over c_{44}^R(q_\perp,q_z)}={1\over c_{44}^0(q_\perp,q_z)}\bigg[1
-{n_0\tilde{\epsilon}_1\over c_{44}^0(q_\perp,q_z)}{n_n(q_\perp,q_z)\over n_0}\bigg],
\end{equation}
where $n_n(q_\perp,q_z)$ has the rather complicated integral expression given in 
Eq. (6.7) below. The corrections to the tilt modulus incorporated in $n_n$ can be
interpreted in terms of a disentangled fraction of the flux liquid - hence
a ``normal-fluid component''. When the nonlocality of the 
intervortex interaction in the field direction is neglected, Eq. (\ref{result})
becomes identical to the result obtained by T\"auber and Nelson (see Eq. (3.33) of Ref. \onlinecite{TN}). In this case the long-wavelength $c_{44}$ 
is not renormalized in infinitely thick samples.
 
Our result, Eq. (\ref{result}), is also simply related to
the Larkin-Vinokur formula given in Eq. (\ref{LVresult}).
This is immediately seen by  introducing a normal
fluid fraction in Eq. (\ref{LVresult}) as
$n_n=n_0-n_s$, and then expanding 
for small values of the normal fluid fraction, $n_n/n_0<< 1$,
to obtain 
\begin{equation}
\label{LVapprox}
{1\over c_{44}^{LV}}\approx{1\over c_{44}^0}\bigg[1
-{n_0\tilde{\epsilon}_1\over c_{44}^0}{n_n\over n_0}\bigg],
\end{equation}
with $c_{44}^0$ given by Eq. (\ref{tilt0}). 
This expression is formally identical to the long-wavelength 
($q_\perp=0,q_z=0$) limit of our result. 

We find that interaction effects in a clean flux liquid do lead to a nonvanishing 
renormalization of the tilt modulus in the thermodynamic limit of thick samples.
This correction is present only if the nonlocality of the intervortex interaction
is properly incorporated. The correction remains, however, small at all but
very low $(B<1{\rm Tesla})$ fields.
Our results are perturbative and cannot be used to infer quantitative conclusions
about the existence of a true disentangled flux liquid phase.
One of the main outcomes of our work is the development of a transparent 
hydrodynamic framework that can be used to study the role of the nonlocality
of the intervortex interaction on the tilt response, both in clean materials
and in the presence of disorder of various geometries. Note that in
conventional, Gaussian hydrodynamics the effect of disorder on $c_{44}$
cannot be detected.
 
In section II we discuss the general form of the London free energy used as
the starting point
to study the magnetic properties of superconductors
in the mixed state. The various response functions of interest are also
defined there. 
After
discussing the response to a tilt field in section III, we review and contrast
in sections IV and V, respectively,
the results obtained by conventional Gaussian hydrodynamics and by the local 
boson mapping. After showing  how hydrodynamics can be derived from the
boson model in section VI, we introduce our non-Gaussian hydrodynamic model and discuss its
relationship to
previous work. Our results are discussed in section VII. Finally, a
rigorous derivation of the nonlocal, non-Gaussian hydrodynamics from the
charged boson analogy is displayed in Appendix A, and the perturbative evaluation of the renormalization of $c_{44}$
from interactions is displayed in Appendix B.

\section{Magnetic Response of the Vortex Array}

High-$T_{c}$ superconductors are uniaxial, strongly type-II materials with very large values
of the Ginzburg-Landau parameter $\kappa=\lambda/\xi$. For applied fields 
${H}_{c1}<<{H}<<{H}_{c2}$, their mixed state 
can be described in the
London limit with a  Ginzburg-Landau Hamiltonian given by
\begin{eqnarray}
\label{HGL}
{\cal H}[\theta,{\bf A}]&=&{1\over 2}\int_{\bf r}\bigg\{
   {c^2\over 4\pi\tilde{\lambda}_\mu^2}\Big({\phi_0\over 2\pi}\partial_\mu\theta -A_\mu\Big)^2+\nonumber\\
& &   +{1\over 4\pi}(\bbox{\nabla}\times{\bf A})^2\bigg\}.
\end{eqnarray}
Here the $z$ direction has been chosen along the anisotropy ($c$) axis of the superconductor.
Greek indices $\mu,\nu,...$ run over all Cartesian components ($\mu=x,y,z$) and summation is
intended in Eq. (\ref{HGL}). Latin indices $i,j,k,...$ run only over $x$ and $y$. 
The integral $\int_{\bf r}...\equiv\int_0^L dz\int d{\bf r}_\perp ...$ is over the 
volume $\Omega=LA$ of the superconductor, with $L$ the thickness in the direction of the 
$c$ axis and $A$ the area in the $ab$ plane.
Also, $\tilde\lambda_\mu =\lambda_\mu / (1-H/ H_{c2})^{1/2}\ $, where 
$\lambda_x =\lambda_y=\lambda_\perp$ are the penetration depths 
from supercurrents in
the $ab$ plane, while $\lambda_z=p\lambda_{\perp}$ is the penetration depth 
from supercurrents along the $c$ axis, with
$p$ the anisotropy ratio  arising from
an effective mass tensor for the superconducting electrons $(p=(m_z/m_{\perp})^{1/2})$.
Finally, ${\bf A}$ is the total vector potential, with ${\bf B}=\bbox{\nabla}\times{\bf A}$
the internal field in the material, and $\phi_0=hc/2e$ is the flux quantum. 
The corresponding Gibbs free energy functional is 
\begin{equation}
\label{GibbsGL}
{\cal G}[\theta,{\bf H}]={\cal H}[\theta,{\bf A}]
  -{1\over 4\pi}\int_{\bf r}{\bf B}\cdot{\bf H},
\end{equation}
where ${\bf H}=\bbox{\nabla}\times{\bf A}^{\rm ext}$ is the applied external field.

The London free energy functional can be rewritten in terms of interacting vortex lines
by introducing a ``vortex line density'' vector defined as
\begin{equation}
\label{densityvector1}
\hat{\bf T}({\bf r})={1\over 2\pi}\bbox{\nabla}\times(\bbox{\nabla}\theta).
\end{equation}
Here and below a hat ($\hat{~}$) is used, when needed, to distinguish microscopic fluctuating 
quantities from average ones.
We will specifically consider situations where the magnetic field responsible for the
onset of the vortex state is applied along the $z$ direction. 
Vortex line configurations are then conveniently characterized by a set
of $N$ single-valued functions ${\bf r}_{n}(z)$, which specify the position of
the $n$-th vortex line in the $xy$ plane as it wanders along the $z$ axis.
The three-dimensional position of each flux line is parametrized as 
${{\bf R}_n}(z)=[{{\bf r}_n}(z),\,z]$ and the vortex density vector can be written as 
\begin{equation}
\label{densityvector2}
\hat{\bf T}({\bf r})=\sum_{n=1}^{N}{d{\bf R}_n(z)\over dz}\delta^{(2)}({\bf r}_\perp
   -{\bf r}_n(z)),
\end{equation}
where ${\bf r}=({\bf r}_\perp,z)$. The vortex density vector can be written as
$\hat{\bf T}({\bf r})=\big(\hat{\bf t},\hat{n}\big)$, where 
$\hat{\bf t}$ is a two-dimensional vector describing the local tilt of flux lines 
away from the direction of the external field and 
$\hat{n}$ is the areal density of vortices, 
\begin{eqnarray}
\label{nhat}
& & \hat{n}({\bf r})=\sum_{n=1}^N\delta^{(2)}({\bf r}_\perp-{\bf r}_n(z)),\\
& & \hat{\bf t}({\bf r}) = \sum_{n=1}^N {d{\bf r}_n(z)\over dz}
\delta^{(2)}({\bf r}_{\perp}-{\bf r}_n(z)) \;.
\label{that}
\end{eqnarray}
The vortex density vector is also directly
related to the superfluid velocity of the electrons in the superconductor, ${\bf v}^s=(\phi_0/2\pi)\bbox{\nabla}\theta-{\bf A}$, by
\begin{equation}
\label{superf}
\phi_0\hat{\bf T}-\hat{\bf B}=\bbox{\nabla}\times{\bf v}^s.
\end{equation}
The Cartesian components of the local supercurrent are 
$j^s_\mu=(c/4\pi\tilde\lambda_\mu^2)v^s_\mu$ (no summation over
$\mu$ intended here).
After some manipulations (see, for instance, Ref. \onlinecite{chenteitelI} for the details)
and neglecting spin wave fluctuations, one obtains
\end{multicols}
\begin{equation}
\label{Gv}
{\cal G}[\hat{\bf T},{\bf H}]={1\over 8\pi\Omega}\sum_{\bf q}\bigg\{
  \big[\phi_0\hat{T}_\mu({\bf q})-\hat{B}_\mu({\bf q})\big] U_{\mu\nu}({\bf q})
   \big[\phi_0\hat{T}_\nu({-\bf q})-\hat{B}_\nu({-\bf q})\big]
   +|\hat{\bf B}({\bf q})|^2 -2{\bf H}({\bf q})\cdot \hat{\bf B}(-{\bf q})\bigg\},
\end{equation}
with 
\begin{eqnarray}
\label{int1}
U_{\mu\nu}({\bf q})={1\over \tilde\lambda_\perp q^2}
     \Big[\delta_{\mu\nu}-\delta_{\mu i}\delta_{\nu j}
      {(\tilde\lambda^2_z-\tilde\lambda^2_\perp)q^2_\perp\over
          \tilde\lambda^2_zq^2_\perp+\tilde\lambda^2_\perp q^2_z} P^T_{ij}({\bf q}_\perp)\Big].
\end{eqnarray}
Here,  ${\bf q}=({\bf q_{\perp}},q_z)$ and 
$\  P_{ij}^T({\bf q}_{\perp}) = \delta_{ij}-\hat{q}_{\perp i}\hat{q}_{{\perp}j}$
is the two-dimensional transverse projection operator, with $\hat{\bf q}_\perp={\bf q}_\perp/q_\perp$.
The corresponding longitudinal projection operator is 
$P^L_{ij}({\bf q}_\perp)=\delta_{ij}-P^T_{ij}({\bf q}_\perp)$.

In this paper we will only consider magnetic field fluctuations due to fluctuations
in the vortices' degrees of freedom. This London part of the field fluctuations is 
obtained by minimizing the 
Ginzburg-Landau free energy (\ref{Gv}) for fixed vortex configurations
$\hat{T}({\bf q})$ and it is given by
\begin{equation}
\label{minind}
\hat{\bf B}({\bf q})=\hat{\bf B}^V({\bf q})+\hat{\bf B}^M({\bf q}),
\end{equation}
where $\hat{\bf B}^V({\bf q})$ is the part of the internal field due to the vortices, 
\begin{eqnarray}
\label{Vinduction}
\hat{B}^V_{\mu}({\bf q})& =  &
  \big({\bf 1}+{\bf U}({\bf q})\big)^{-1}_{\mu\sigma}U_{\sigma\nu}({\bf q})\phi_0\hat{T}_\nu({\bf q})\\\nonumber
  &=& {1\over 1+\tilde\lambda^2_\perp q^2}
   \bigg[\delta_{\mu\nu}
   -\delta_{\mu i}\delta_{\nu j}{(\tilde\lambda_z^2-\tilde\lambda_\perp^2)q_\perp^2\over
   1+\tilde\lambda^2_\perp q_z^2+\tilde\lambda_z^2 q_\perp^2}P^T_{ij}({\bf q}_\perp)\bigg]
   \phi_0\hat{T}_\nu({\bf q}),
\end{eqnarray}
and $\hat{\bf B}^M({\bf q})$ is the Meissner
response of the material to a spatially inhomogeneous external field, 
\begin{eqnarray}
\label{Minduction}
\hat{B}^M_{\mu}({\bf q})&=  &
     \big({\bf 1}+{\bf U}({\bf q})\big)^{-1}_{\mu\nu}H_\nu({\bf q})\\\nonumber
 & =& {1\over 1+\tilde\lambda^2_\perp q^2}
   \bigg[\tilde\lambda_\perp^2 q^2\delta_{\mu\nu}
   +\delta_{\mu i}\delta_{\nu j}{(\tilde\lambda_z^2-\tilde\lambda_\perp^2)q_\perp^2\over
   1+\tilde\lambda^2_\perp q_z^2+\tilde\lambda_z^2 q_\perp^2}P^T_{ij}({\bf q}_\perp)\bigg]
   H_\nu({\bf q}).
\end{eqnarray}
In addition to the contributions given in Eq. (\ref{minind}), there are field fluctuations
representing thermal deviations from
the solution of the London equation, which are neglected here.
By inserting Eqs. (\ref{Vinduction}) and (\ref{Minduction}) into Eq. (\ref{Gv}), we obtain the 
vortex free energy functional expressed entirely in terms of vortex degrees of freedom,
\begin{equation}
\label{gibbsth}
{\cal G}[\hat{\bf T},{\bf H}]={1\over 2\Omega}\sum_{\bf q}\bigg\{
       \hat{T}_\mu({\bf q})V_{\mu\nu}({\bf q}) \hat{T}_\nu(-{\bf q})
       -{1\over\phi_0}H_\mu({\bf q})V_{\mu\nu}({\bf q}) \hat{T}_\nu(-{\bf q})
       -{1\over 4\pi}H_\mu({\bf q})\big({\bf 1}
       +{\bf U}({\bf q})\big)^{-1}_{\mu\nu}H_\nu(-{\bf q})\bigg\},
\end{equation}
where 
\begin{eqnarray}
\label{Vint}
V_{\mu\nu} ({\bf q}) &= &
    V_0\big({\bf 1}+{\bf U}({\bf q})\big)^{-1}_{\mu\sigma}U_{\sigma\nu}({\bf q})\\\nonumber
  & =&\frac{V_0}{1 +\tilde  \lambda_{\perp}^2 q^2}
   \bigg[\delta_{\mu\nu} 
  - \delta_{\mu i}\delta_{\nu j}
   \frac{ (\tilde  \lambda_z ^2 - \tilde \lambda_{\perp}^2)
q_{\perp}^2}{1 + \tilde \lambda_{\perp}^2 q_z^2 + \tilde  \lambda_z ^2
q_{\perp}^2} P_{ij}^T({\bf q}_{\perp})\bigg],
\end{eqnarray}
are the Fourier components of the anisotropic intervortex interaction, 
with $V_0=\phi_0^2/4\pi$. 
One important property of the intervortex interaction is its nonlocality.
In particular, the nonlocality in the $z$ direction, reflecting that 
flux-line elements at different $z$ heights repel each other via a 
Yukawa-like potential, will play a very important role in the discussion below.

\begin{multicols}{2}
The Gibbs free energy of the vortex system is given by 
\begin{equation}
G({\bf H},T)=-k_BT\ln{\cal Z}({\bf H},T),
\end{equation}
where  
\begin{equation}
\label{partf}
{\cal Z}({\bf H},T)=\int '{\cal D}\hat{\bf T}({\bf r})e^{-{\cal G}/k_BT}
\end{equation}
is the canonical partition function. The prime over the integral sign indicates
that the integration must be performed with the constraint $\bbox{\nabla}\cdot\hat{\bf B}=0$.
The average local field in the superconductor is then given by 
\begin{equation}
\label{meanB}
{\bf B}({\bf r})=<\hat{\bf B}({\bf r})>=-4\pi{\delta G\over\delta{\bf H}({\bf r})},
\end{equation}
where the brackets denote a statistical average with Boltzmann weight $\sim\exp[-{\cal G}/k_BT]$.

For a spatially homogeneous external field applied along the $z$
direction, ${\bf H}({\bf r})=\hat{\bf z}H_0$, we obtain the familiar form 
\cite{blatter},
\begin{eqnarray}
\label{gibbs0}
{\cal G}_0(\hat{\bf T},H_0) = & &   - NL\frac{H_0 \phi_0}{4\pi} \\ \nonumber
  & & +\frac{1}{2\Omega} \sum_{\bf q}\hat{T}_\mu({\bf q})V_{\mu\nu}({\bf q}) \hat{T}_\nu(-{\bf q}).
\end{eqnarray}
For a uniform applied field  ${\bf H}=\hat{\bf z}H_0$, the Meissner part of the
transverse local field given in Eq. (\ref{Minduction}) vanishes.
The local field in the superconductor is entirely due to the vortices
and it given by Eq. (\ref{Vinduction}).
From here on we will always refer to the vortex system created
by the homogeneous field ${\bf H}=\hat{\bf z}H_0$ and the local
field is to be understood as the field given by Eq.
(\ref{Vinduction}).

The focus of this paper is on the response of the vortex array created by the 
external field $\hat{\bf z}H_0$ to a small additional
spatially inhomogeneous external field $\delta{\bf H}({\bf r})$. The Gibbs free
energy functional in the presence of this perturbation can be written as
\begin{equation}
\label{gibbs}
{\cal G}(\hat{\bf T},\hat{\bf z}H_0+\delta{\bf H})=
  {\cal G}_0(\hat{\bf T},H_0) +
  \delta {\cal G}(\hat{\bf T},\delta {\bf H}),
\end{equation}
where ${\cal G}_0$ is given by Eq. (\ref{gibbs0}) and the perturbation is
\begin{eqnarray}
\label{gibbspert}
\delta {\cal G}(\hat{\bf T},\delta {\bf H})&=&-{1\over 4\pi}\int_{\bf r}\hat{\bf B}^V
  \cdot\delta{\bf H}\\
&=& -{1\over c}\int_{\bf r}\hat{\bf j}^s\cdot\delta{\bf A}^{\rm ext}.
\end{eqnarray}
The local field $\hat{\bf B}^V$ in Eq. (\ref{gibbspert}) is the field 
in the absence of the perturbation $\delta{\bf H}$ and is related
to the vortex degrees of freedom via Eq. (\ref{Vinduction}). It does not include
the Meissner response to the perturbation $\delta{\bf H}$.
The supercurrent is defined as $\hat{\bf j}^s=(c/4\pi)\bbox{\nabla}\times\hat{\bf B}^V$.

Below we will use $\langle ...\rangle_0$ to denote a statistical average over the unperturbed ensemble
described by ${\cal G}_0$, while $\langle ...\rangle_{H}$ will denote the average over the
perturbed ensemble, with free energy given by Eq. (\ref{gibbs}).
The mean local field ${\bf B}^H$ in the material in the presence of the perturbation
$\delta{\bf H}$ can be written as the sum of vortex and Meissner
parts as
\begin{equation}
\label{response}
{\bf B}^H({\bf q})=\langle\hat{\bf B}^V({\bf q})\rangle_H+ 
    \delta{\bf B}^M({\bf q}),
\end{equation}
where $\delta{\bf B}^M({\bf q})$ is the Meissner response to the perturbation, given by
Eq. (\ref{Minduction}) with ${\bf H}({\bf q})=\delta{\bf H}({\bf q})$.
To linear order in the perturbing field, the vortex contribution can be expressed in terms 
of correlation functions in the unperturbed ensemble as,
\begin{equation}
\label{vortexpert}
\langle\hat{B}_\mu^V({\bf q})\rangle_H = \langle\hat{B}^V_\mu({\bf q})\rangle_0+
 {\beta\over 4\pi}\langle\hat{B}^V_\mu({\bf q})\hat{B}^V_\nu(-{\bf q})\rangle_0^c\delta H_\nu({\bf q}),
\end{equation}
where $\langle ...\rangle^c$ is the connected part of the correlator, i.e.,
$\langle AB\rangle^c=\langle AB\rangle-\langle A\rangle\langle B\rangle$.
Finally, the corresponding linear response function defines the magnetic susceptibility
$\chi_{ij}({\bf q})$ of the material according to
\begin{equation}
\label{linresp}
B_{\mu}^H({\bf q})-\langle\hat{B}^V_\mu({\bf q})\rangle_0=
\big[4\pi\chi_{\mu \nu}({\bf q})+ 
\delta_{\mu \nu}\big]\delta H_{\nu}({\bf q}).
\end{equation}
The components of the susceptibility tensor can also be expressed in terms of vortex 
density correlations,
\begin{eqnarray}
\label{chi}
4\pi\chi_{\mu \nu}({\bf q})=-{\frac{V_{\mu \nu}}{V_0}} +  
{\frac{{\phi}_{0}^2}
{k_B T V_{0}^2}} V_{\mu \sigma}({\bf q})V_{\nu \lambda}(-{\bf q})T_{\sigma \lambda}({\bf q}),
\end{eqnarray}
where $T_{\mu \nu}({\bf q})$ is the correlation function of the vortex density vector,
\begin{equation}
\label{denscorr}
T_{\mu \nu}({\bf q})=\langle\hat{T}_{\mu}({\bf q})\hat{T}_{\nu}(-{\bf q})\rangle_0^c.
\end{equation}
The density-vector correlation function can be expressed
in terms of derivatives of the partition function
of the perturbed system as
\begin{eqnarray}
\label{tiltcorrder}
\langle\hat{T}_\mu({\bf q})\hat{T}_\nu({\bf q}')\rangle_0^c= & &
(\phi_0k_BT)^2(V^{-1})_{\mu \kappa}(V^{-1})_{\nu \lambda}\\ \nonumber
& & \times\Big[{\delta^2\ln{\cal Z}(H_0\hat{\bf z}+\delta{\bf H},T)\over
\delta H_\kappa({\bf q})\delta H_\lambda({\bf q'})}\Big]_{\delta{\bf H}=0},
\end{eqnarray}
where $(V^{-1})_{\mu \nu}$ are the components of the inverse of the interaction
tensor (\ref{Vint}).

The tensor $T_{\mu\nu}$ is block diagonal, with
$T_{\mu\nu}=(T_{ij},T_{zz})$.
The component $T_{zz}$ is the density-density
correlation function or structure function of the vortices,
\begin{equation}
T_{zz}({\bf q})=S({\bf q})=\langle\delta\hat{n}({\bf q})\delta\hat{n}(-{\bf q})\rangle_0,
\end{equation}
where $\delta\hat{n}({\bf q})=\hat{n}({\bf q})-n_0\Omega\delta_{{\bf q,0}}$ describes the fluctuation
of the local density field from its mean value $n_0=B_{0z}/\phi_0$,
with $B_{0z}\approx H_0$ the equilibrium value of the $z$ component of
the internal field.
The in-plane part $T_{ij}$ is the tilt-tilt autocorrelator and it is the central 
quantity of interest  here. It can be written in terms of transverse and 
longitudinal components as 
\begin{equation}
\label{tiltauto}
T_{ij}({\bf q})=T_L({\bf q})P^L_{ij}({\bf q}_\perp)+T_T({\bf q})P^T_{ij}({\bf q}_\perp).
\end{equation}
The transverse part of the tilt autocorrelator determines the tilt modulus
of the vortex array. The wave-vector-dependent tilt modulus is defined by
\begin{equation}
\label{tmod}
T_T ({\bf q})=\frac{n_0^2 {k_B} T}{c_{44}(q_\perp,q_z)}.
\end{equation}

Finally, in order to make contact with the literature, it is useful
to write the perturbing field in terms of a vector potential,
$\delta{\bf H}=\bbox{\nabla}\times\delta{\bf A}^{\rm ext}$.
The linear response to the vector potential $\delta{\bf A}^{\rm ext}$
is then characterized by the helicity tensor ${\Upsilon}_{\mu\nu}$, 
which relates the induced current to $\delta{\bf A}^{\rm ext}$,
\begin{equation}
\label{indcurrent}
j^H_\mu({\bf q})=-c{\Upsilon}_{\mu\nu}({\bf q})\delta A_\nu^{\rm ext}({\bf q}),
\end{equation}
where ${\bf j}^H$ is the total screening current induced in the material by the
perturbing vector potential, comprising of both the vortex and Meissner contributions.
The helicity tensor can be immediately related to the components of the susceptibility
tensor,
\begin{equation}
{\Upsilon}_{\mu\nu}({\bf q}_\perp)=-\epsilon_{\mu\sigma\xi}\epsilon_{\nu\alpha\beta}
q_\sigma q_\alpha \chi_{\xi\beta}({\bf q}).
\end{equation}
Using Eq. (\ref{chi}), it can also be expressed in term of the correlations of the vortex density
tensor.

\section{Tilting field}

In the remainder of this paper we focus on
the response of the vortex array to a spatially inhomogeneous field 
$\delta{\bf H}_\perp({\bf q})$ applied normal to the direction of $H_0$ 
and that tilts the flux lines away from the $z$ direction.
As discussed by Chen and Teitel \cite{chenteitelI}, we distinguish two types of perturbations.
The first is a tilt perturbation, corresponding to a tilting field
which is spatially homogeneous in the $xy$ plane and may be modulated in
the $z$ direction. The long wavelength response
to this tilt perturbation is 
determined by the long wavelength tilt modulus, $c_{44}$, defined as
\begin{equation}
\label{longtilt}
\frac{n_0^2 {k_B} T}{c_{44}}=\lim_{q_z\rightarrow
0}\lim_{q_\perp\rightarrow 0}T_T(q_{\perp}, q_z).
\end{equation}
The order of the limits ($q_{\perp}\rightarrow 0$ first, followed
by $q_z \rightarrow 0$) is important here and  reflects the physical situation of the
relevant experiment. The vanishing of the long wavelength tilt modulus
signals the onset of a transverse Meissner
effect, where the perturbing field is
completely expelled from the material (as seen from Eq. (\ref{gibbsth}), the corresponding 
static susceptibility equals $ -1/4\pi $). This occurs, for instance, in vortex arrays 
pinned by columnar defects.

The second physical experiment of interest here is the response to a tilting
field $\delta{\bf H}_\perp({\bf q}_\perp)$ which is spatially homogeneous in the
$z$ direction (i.e., independent of $q_z$) and generates a 
shear perturbation of the vortex array. Such a field can be obtained
from a vector potential $\delta{\bf A}^{\rm ext}=\hat{\bf z}\delta A_z^{\rm ext}({\bf r}_\perp)$,
which induces screening currents along the $z$ direction. In the literature 
the response of the superconductor
to such a shear perturbation is often characterized by the corresponding 
component of the helicity modulus (${\Upsilon}_{zz}({\bf q}_\perp)$)
defined in Eq. (\ref{indcurrent}), which in turn is
related to the transverse part of the tilt-tilt correlator by
\begin{equation}
\label{helicity}
{\Upsilon}_{zz}(q_\perp)={1\over 4\pi}{q_\perp^2\over 1+q_\perp^2\tilde\lambda_z^2}
  \bigg[1-{V_0\over k_BT}{T_T(q_\perp, q_z=0)\over 1+q_\perp^2\tilde\lambda_z^2}\bigg],
\end{equation}
where the first term arises from the Meissner part of the response.
The long wavelength limit of the helicity modulus is
\begin{equation}
\lim_{q_\perp\rightarrow 0}{\Upsilon}_{zz}(q_\perp)={q_\perp^2\over 4\pi}
  \Big[1-{V_0\over k_BT}\lim_{q_\perp\rightarrow 0}T_T(q_\perp, q_z=0)\Big].
\end{equation}
The vanishing of $\lim_{q_\perp\rightarrow 0}{T_T}(q_\perp, {q_z}=0)$ yields
$\lim_{q_\perp\rightarrow 0}4\pi{\Upsilon}_{zz}(q_\perp)/q^2_\perp =1$,
which corresponds to a perfect Meissner response in the $z$ direction and signals 
longitudinal superconductivity.

We emphasize, however, that both the perturbations just described simply
probe the magnetic response of the superconductor, which is the
true equilibrium test of superconductivity. 
In fact the relevant response function in each case (tilt or helicity modulus)
is simply  the transverse
part of the susceptibility tensor,
\begin{equation}
\chi_T({\bf q})=P^T_{ij}({\bf q}_\perp)\chi_{ij}({\bf q}).
\end{equation}
The long wavelength tilt modulus is given by
\begin{equation}
{n_0^2V_0\over c_{44}}=1+4\pi\lim_{q_z\rightarrow 0}\chi_T(q_\perp=0,q_z),
\end{equation}
and the component of the helicity modulus that controls longitudinal superconductivity
is
\begin{equation}
\lim_{q_\perp\rightarrow 0}{\Upsilon}_{zz}(q_\perp)=-\lim_{q_\perp\rightarrow 0}{q_\perp}^2\chi_T(q_\perp, q_z=0).
\end{equation}

In a flux-line {\it lattice} the transverse part of the tilt-tilt correlator 
is non-analytic at small wave-vectors
and the different order of limits of the two perturbations discussed above is important.
This is because the vortex array has a nonzero long wavelength shear
modulus, $c_{66}$. As a 
result, the flux lattice exhibits longitudinal superconductivity, with
$\lim_{q_\perp\rightarrow 0}T_T(q_\perp, q_z=0)=0$, and
\begin{equation}
\lim_{q_\perp\rightarrow 0}\chi^{\rm lattice}_T(q_\perp, q_z=0)=-{1\over 4\pi},
\end{equation}
but no transverse Meissner effect, as
$\lim_{q_z\rightarrow 0}T_T({\bf q}_\perp=0, q_z)\not=0$ and
\begin{equation}
\lim_{q_z\rightarrow 0}\chi^{\rm lattice}_T(q_\perp=0, q_z)=-{1\over 4\pi}+{V_0n_0^2\over c_{44}}.
\end{equation}

In a flux-line {\it liquid}, in contrast, we find that the order of limits is not important and the flux array
in general exhibits neither longitudinal superconductivity, nor perfect Meissner effect, as
\begin{eqnarray}
\lim_{q_z\rightarrow 0}\chi^{\rm liquid}_T& &(q_\perp=0, q_z)
=\lim_{q_\perp\rightarrow 0}\chi^{\rm liquid}_T(q_\perp, q_z=0)\\\nonumber
 & &= -{1\over 4\pi}+{V_0n_0^2\over c^R_{44}},
\end{eqnarray}
where $c_{44}^R$ is the flux liquid tilt modulus, renormalized by interaction effects.
We will see below, however, that interactions can yield a strong upward renormalization 
of $c_{44}$ even in clean flux liquids.

\section{Gaussian Hydrodynamics}

A useful framework for discussing the long wavelength properties
of flux-line liquids that 
naturally incorporates all nonlocalities of the intervortex interaction
is hydrodynamics, where
vortex fluctuations are described in terms of a few coarse-grained fields.
By long wavelengths, we mean wavelengths large compared to the
spacing between ${\rm CuO}_2$ planes in the $\hat{\bf z}$ direction, and large
compared to the intervortex spacing in the $ab$ plane normal to $\hat{\bf z}$.

The coarse-grained hydrodynamic fields for a flux-line liquid are the fluctuating
areal density,
\begin{equation}
\label{density}
\hat{n}^H({\bf r})=\sum_{n=1}^N\delta_{BZ}^{(2)}({\bf r}_\perp-{\bf r}_n(z)),
\end{equation}
and a tilt field,
\begin{equation}
\label{tiltfield}
\hat{\bf t}^H({\bf r}) = \sum_{n=1}^N {\frac{d {\bf r}_n}{d z}}
\delta_{BZ}^{(2)}({\bf r}_{\perp}-{\bf r}_n(z)) \;.
\end{equation}
Here $\delta_{BZ}^{(2)}({\bf r}_{\perp})$ is a smeared-out
two-dimensional $\delta$-function with a finite spatial extent of the order of the inverse
of the Brillouin zone boundary $k_{BZ}=\sqrt{4\pi n_0}$.
It is defined as
\begin{equation}
\delta_{BZ}^{(2)}({\bf r}_{\perp})={1\over A}\sum_{q_\perp\leq k_{BZ}}
  e^{-i{\bf q}_\perp\cdot{\bf r}_\perp}.
\end{equation}
We stress that these hydrodynamic fields differ from the microscopic fields defined
in Eq. (\ref{nhat}) and (\ref{that}) as they are coarse-grained quantities
obtained by averaging out the more microscopic and rapidly varying degrees of freedom.

A {\it Gaussian} hydrodynamic free energy containing terms quadratic in the deviations of the
fields from their equilibrium values can be obtained by coarse-graining
the  microscopic energy of interacting
vortices given in Eq. (\ref{gibbs}), with the result \cite{mcment},
\begin{eqnarray}
\label{linhydror}
F_G={1\over 2n^2_0}\int_{\bf r}\int_{{\bf r}'} & & \Big[B({\bf r}-{\bf r}')\delta\hat{n}^H({\bf r})
  \delta\hat{n}^H({\bf r}')\\ \nonumber
 & &  +K({\bf r}-{\bf r}')\hat{\bf t}^H({\bf r})\cdot \hat{\bf t}^H({\bf r}')\Big],
\end{eqnarray}
where $\delta\hat{n}^H({\bf r})=\hat{n}^H({\bf r})-n_0$ and $B({\bf r})$ and $K({\bf r})$
are nonlocal liquid elastic constants. 
The density and tilt fields are not independent quantities, but are related by a ``continuity''
equation expressing the constraint that  vortex lines cannot start or stop inside
the sample,
\begin{equation}
\label{constraint}
\partial_z\delta\hat{n}^H+{\bbox{\nabla}}_\perp\cdot\hat{\bf t}^H=0.
\end{equation}
The Gaussian hydrodynamic free energy is rewritten
in a more familiar form by passing to Fourier space, 
\begin{equation}
\label{linhydro}
F_G={1\over 2n^2_0\Omega}\sum_{\bf q}\Big[c_{11}^0({\bf q})|\delta\hat{n}^H({\bf q})|^2
+c_{44}^0({\bf q})|\hat{\bf t}^H({\bf q})|^2\Big],
\end{equation}
where $c_{11}^0({\bf q})$ and $c_{44}^0({\bf q})$ are the bare compressional and tilt moduli
of the flux liquid. The compressional modulus is given by
\begin{eqnarray}
\label{barecomp}
 c_{11}^0({\bf q})=   {B_{0z}^2\over 4\pi} \frac{1+q^2
\tilde{\lambda}_{\perp}^2 p^2}{ (1 + q^2 \tilde{\lambda}_{\perp}^2 )(1+q_z^2
 \tilde{\lambda}_{\perp}^2 + q_{\perp}^2 p^2
\tilde{\lambda}_{\perp}^2 )}                 \;  . 
\end{eqnarray}
The bare tilt modulus is found to be to a good approximation identical to 
the flux lattice tilt modulus given in
Eqs. (\ref{tilt}-\ref{tiltc}) \cite{mcmdis,mcment}.

In this Gaussian approximation, the probability of a fluctuation is
proportional to $\exp(-F_G/k_BT)$ and averages must  be
carried out subject to the continuity constraint, Eq. (\ref{constraint}).
The correlation functions of the hydrodynamic fields are then immediately
calculated and are given by
\begin{equation}
\label{elcorII}
\langle \delta \hat{n}^H (-{\bf q}) \delta \hat{n}^H ({\bf q}) \rangle_{G}  = \frac {n_{0}^2k_B T q_{\perp}^2}{c_{44}^0({\bf q}) q_z^2 + c_{11}^0({\bf
q}) q_{\perp}^2 } \; ,
\end{equation}
\begin{equation}
\label{elcorIII}
\langle \hat{t}^H_i (-{\bf q}) \delta \hat{n}^H ({\bf q}) \rangle_{G} = \frac{n_{0}^2k_B T q_{\perp i} q_z}{c_{44}^0({\bf q}) q_z^2 + c_{11}^0({\bf
q}) q_{\perp}^2 } \;,
\end{equation}
\begin{equation}
\label{elcorI}
\langle \hat{t}^H_i (-{\bf q}) \hat{t}^H_j ({\bf q}) \rangle_{G} = T^0_T ({\bf q}) P_{ij}^T({\bf q}_{\perp}) +
T^0_L({\bf q}) P_{ij}^L({\bf q}_{\perp}) \; ,
\end{equation}
with
\begin{equation}
\label{transversetilt}
T^0_T ({\bf q}) = \frac{n_{0}^2k_B T}{c_{44}^0({\bf q})}
\end{equation}
and
\begin{equation}
\label{longittilt}
T^0_L({\bf q}) = \frac{n_{0}^2 k_B T q_z^2}{c_{44}^0({\bf q}) q_z^2 + c_{11}^0({\bf
q}) q_{\perp}^2 }.
\end{equation}
The long wavelength tilt modulus is determined by
the transverse part of the tilt autocorrelator, according to 
Eq. (\ref{transversetilt}). To this Gaussian order it is then identically given by its bare value, $c_{44}^0$, given in Eq. (\ref{tilt0}).
Gaussian hydrodynamics does not allow for any renormalization 
of the tilt modulus, even in the presence of disorder. This is because
a disorder potential couples to the flux-line areal density that, within a
Gaussian theory, is in turn decoupled from the transverse part of the tilt
field. In particular, this naive hydrodynamic theory does not describe
the possibility of a disentangled flux-line liquid, with a 
tilt modulus enhanced by interaction or disorder.
In other words, Gaussian 
hydrodynamics is by definition a theory of
{\it entangled} flux-line liquids.

%%%%%%%%%%%%%%%%%%%%%%%%%%%%%%%%%%%%%%%%%%%%%%%%%%%%%%%%%%%%%%%%%%%%%%%%%%%%%%%%%%%%%%%%%

\section{2d Boson Model}

Considerable progress in understanding the properties of vortex-line 
arrays has been made by employing
the formal analogy between the classical statistical mechanics of directed lines in three
dimensions and the quantum statistical mechanics of two-dimensional bosons. 
The advantage of this approach is that it can incorporate interaction effects
accounting for localization or disentanglement of the vortices.
The drawback is that this model, at least in its simplest implementation
employed by Nelson and coworkers \cite{nelson,seung,drn_ledoussal,TN}, neglects the nonlocality of the
intervortex interaction. We will show below that the nonlocality of the
interaction in the field ($z$) direction plays a crucial role in
controlling the tilt modulus.

In this section we briefly review the local version of the
boson mapping employed by Nelson and coworkers \cite{nelson,seung,drn_ledoussal}
and the results obtained recently for the tilt modulus  by T\"auber and Nelson 
 \cite{TN} using this model. 

Neglecting the nonlocality of the intervortex interaction, the free energy of 
interacting vortex lines in a field ${\bf H}=H_0\hat{\bf z}+\delta{\bf H}_\perp$
given in Eq. (\ref{gibbs}) is approximated as
\end{multicols}
\begin{eqnarray}
\label{freeB}
{\cal G}(\{{\bf r}_n\},{\bf H})= & &NL\big(H_0{\phi_0\over 4\pi}-\epsilon_1\big)
+\int_z \bigg\{\sum_{n=1}^N{\tilde{\epsilon}_1\over 2}\Big[{d{\bf r}_n\over dz}\Big]^2
+{1\over 2}\sum_{m\not= n}V_\perp(|{\bf r}_n(z)-{\bf r}_m(z)|)\bigg\}\\ \nonumber
& & -{\phi_0\over 4\pi}\int_z\sum_{n=1}^N \delta{\bf H}_{\perp}({\bf r}_n(z),z)\cdot{d{\bf r}_n\over dz},
\end{eqnarray}
\begin{multicols}{2} 
where $\tilde{\epsilon}_1=\epsilon_1/p^2$, with 
$\epsilon_1=\epsilon_0\ln\kappa$ the effective line tension and
$\epsilon_0=(\phi_0/4\pi\tilde{\lambda}_\perp)^2$ a characteristic 
energy scale.
The nonlocality relating fields and vortex variables has  been neglected also in the last
term of Eq. (\ref{freeB}).
Two crucial approximations have been made in rewriting the general intervortex energy given in 
(\ref{Gv}) in the form  (\ref{freeB}). First, the leading elastic term in the self-energy part of Eq.
(\ref{Gv}) has been linearized, according to 
$\sqrt{1+{1\over p^2}\big({d{\bf r}_n\over dz}\big)^2}\approx 1+
{1\over 2p^2}\big({d{\bf r}_n\over dz}\big)^2$. 
Secondly, the pair interaction among different flux lines has been 
replaced by an interaction 
$V_\perp(r_\perp)$ acting locally in each constant-$z$ plane, given by 
\begin{equation}
V_\perp(r_\perp)={\phi_0^2\over 8\pi^2\tilde{\lambda}_\perp^2} 
K_0(r_\perp/\tilde{\lambda}_\perp) \; ,
\label{pairint}
\end{equation}
with $K_0(x)$ a modified Bessel function. Of these approximations  the latter is the most severe,
since it amounts to neglecting the $q_z$ dependence of the elastic
constants - an approximation that strongly affects the tilt modulus, as we will see below.
Letting ${\cal G}(\{{\bf r}_n(z)\},{\bf H})=\mu NL+{\cal F}_N(\{{\bf r}_n(z)\},{\bf H})$, with 
$\mu=H_0{\phi_0\over 4\pi}-\epsilon_1=\phi_0(H_0-H_{c1})/4\pi$ a
chemical potential,
the grand canonical partition function of the vortex liquid can be written as
\begin{equation}
{\cal Z}_{\rm gr}({\bf H})=\sum_{N=0}^\infty {1\over N!}e^{\beta L\mu N}\prod_{n=0}^N
\int {\cal D}{\bf r}_n(z) e^{- {\cal F}_N({\bf H})/{k_B T}}.
\label{grandZ}
\end{equation}
The integral in Eq. (\ref{grandZ}) is over all vortex line 
configurations. It has the form of a quantum-mechanical partition function
in the path integral representation for the world lines of $N$ particles of mass
$\tilde{\epsilon}_1$, moving through imaginary time $z$ and
interacting with the repulsive pair potential $V_\perp(r_\perp)$. 
The vortex model with this simplified interaction can
therefore  be mapped into a model of $2D$ massive bosons with instantaneous pairwise
interaction. The mapping results in the following
correspondences: 
\begin{eqnarray}
\label{bosonmap}
& & z\leftrightarrow \tau  \\\nonumber
& & L\leftrightarrow \hbar\beta_{\rm boson}\\\nonumber
& & \tilde{\epsilon}_1\leftrightarrow m\\\nonumber
& & k_BT\leftrightarrow \hbar\\\nonumber
& & H_0{\phi_0\over\ {4 \pi}}-{\epsilon_1}\leftrightarrow \mu,
\end{eqnarray}
where $\beta_{\rm boson}=1/k_BT_{\rm boson}$ is the inverse temperature of the bosons.
The precise mapping of the grand canonical
vortex line partition function  onto the Feynman path integral in imaginary time $\tau$ of 
a gas of two-dimensional bosons 
requires the introduction of a second quantized Hamiltonian 
corresponding to Eq.
(\ref{freeB}) and is described in the literature 
\cite{seung,drn_ledoussal,Feynman,negele}. Some care must be taken 
in dealing with the tilting field $\delta{\bf H}_\perp$ which introduces velocity-dependent terms into
the fictitious boson Lagrangian. 
One important difference between the flux-line array and the boson system is in
the boundary conditions in the fictitious time variable $z$.
The mapping of the free energy (\ref{freeB}) of vortex lines onto the ``action'' of 
two-dimensional bosons is exact only when one imposes periodic boundary conditions for the
flux lines in the $z$ direction, i.e., ${\bf r}_n(L)={\bf r}_n(0)$. In contrast the
natural boundary condition for flux line would be free boundary conditions,
corresponding to $\big({d{\bf r}_n\over dz}\big)_{z=L}=\big({d{\bf r}_n\over dz}\big)_{z=0}=0$. 
As shown by T\"auber and Nelson \cite{TN},
the choice of the boundary conditions does affect the tilt modulus of a finite-thickness sample.
We will not, however, discuss this here as we are ultimately interested in 
infinitely thick samples. 

To complete the mapping, the grand
canonical partition function (\ref{freeB}) is first rewritten in a coherent-state path integral representation 
as
\begin{equation}
{\cal Z}_{\rm gr}({\bf H})=\int{\cal D}\psi({\bf r}_\perp,z)\int{\cal D}\psi^*({\bf r}_\perp,z)e^{-{\cal S}[\psi,\psi^*;{\bf h}]/k_BT}.
\label{path}
\end{equation}
The boson ``action'' in the imaginary-time path 
integral is
\begin{eqnarray}
\label{action}
{\cal S}[\psi,\psi^*& &;{\bf h}]=  \int_{\bf r} \Bigg[\psi^*\Big(k_BT\partial_z 
-{(k_BT)^2\over 2\tilde{\epsilon}_1}\nabla^2_\perp\Big)\psi\\ \nonumber
& & - {k_BT\over 2\tilde{\epsilon}_1}{\bf h}\cdot(\psi^*{\bbox \nabla}_\perp\psi-\psi{\bbox \nabla}_\perp\psi^*)
-{1\over 2\tilde{\epsilon}_1}h^2|\psi|^2\\ \nonumber
& & + \int d{\bf r'}_\perp V_\perp({\bf r}_\perp-{\bf r'}_\perp)|\psi({\bf r}_\perp,z)|^2|\psi({\bf r'}_\perp,z)|^2 \Bigg],
\end{eqnarray}
and ${\bf h}({\bf r})=(\phi_0/4\pi)\delta{\bf H}_\perp({\bf r})$.
The complex fields $\psi$ and $\psi^*$ correspond
to boson annihilation and creation operators in the second quantized Hamiltonian.
It is convenient to rewrite these fields in terms of an amplitude and a phase as
\begin{equation}
\psi({\bf r}_\perp,z)=\sqrt{\hat{n}({\bf r}_\perp,z)}e^{i\theta({\bf r}_\perp,z)}.
\label{amplphase}
\end{equation}
The magnitude $ \hat{n}({\bf r}_\perp,z)$ of the field $\psi$ corresponds to the fluctuating local boson
density.
The phase field $\theta$ determines the boson momentum density,
\begin{equation}
\label{current}
{\bf g}({\bf r}_\perp,z)= k_B T \hat{n} {\bbox \nabla}_{\perp}\theta \; .
\end{equation}
Upon inserting Eq. (\ref{amplphase}) into Eq. (\ref{action}), the action can be written
in terms of density and phase variables as
\begin{eqnarray}
\label{actionntheta}
{\cal S}[\psi,\psi^*;{\bf h}]= & & \int_{\bf r} \Bigg\{ 
ik_BT\hat{n}\partial_z\theta \\ \nonumber
& & +{(k_BT)^2\over 8\tilde{\epsilon}_1}{({\bbox \nabla}_\perp \hat{n})^2\over \hat{n}}
   +{(k_BT)^2\over 2\tilde{\epsilon}_1}\hat{n}({\bbox \nabla}_\perp\theta)^2 \\ \nonumber
& & -{k_BT\over \tilde{\epsilon}_1}i\hat{n}{\bf h}\cdot{\bbox \nabla}_\perp\theta
    -{h^2\over 2\tilde{\epsilon}_1}\hat{n}\\ \nonumber
& & + \int d{\bf r'}_\perp V_\perp({\bf r}_\perp-{\bf r'}_\perp)\hat{n}({\bf r}_\perp,z)\hat{n}({\bf r'}_\perp,z)\Bigg\},
\label{nonlinearS}
\end{eqnarray}
where we have dropped surface terms that vanish for periodic boundary conditions.

The tilt-tilt correlator $T_{ij}({\bf q})$ can be calculated using
Eq. (\ref{tiltcorrder}), with the result,

\end{multicols}
\begin{eqnarray}
\label{TNresult}
T_{ij}({\bf q}) =
\frac{k_B T}{\tilde{\epsilon}_1}  \delta_{ij}\langle \hat{n}({\bf q})\rangle_{{\bf h}=0} 
- \; \Big(\frac{k_B T}{\Omega{\tilde{\epsilon}_1}}\Big)^2 \sum_{{\bf p}, {\bf p}'}
p_{{\perp}i} p'_{{\perp}j}  \langle \hat{n}({\bf q}-{\bf p}) \theta ({\bf p})
\hat{n}(-{\bf q} - {\bf p}') \theta ({\bf p}') \rangle_{{\bf h}=0} \;,
\end{eqnarray}
\begin{multicols}{2}
where the brackets denote an average over the full nonlinear action (\ref{actionntheta}),
evaluated at ${\bf h}=0$.

To proceed, a standard approximation is to consider only small fluctuations of the fields
from their mean values. Letting
\begin{equation}
\hat{n}({\bf r}_\perp,z)=n_0+\delta \hat{n}({\bf r}_\perp,z),
\end{equation}
and retaining only terms quadratic in the fields in the action, the corresponding Gaussian action in 
zero tilting field is given by
\begin{eqnarray}
\label{linearS}
{\cal S}_G[\psi,\psi^*;{\bf 0}]= & & \int_{\bf r} \Bigg\{ 
ik_BT\delta \hat{n}\partial_z\theta \\ \nonumber
& & +{(k_BT)^2\over 8\tilde{\epsilon}_1}{({\bbox \nabla}_\perp \delta
\hat{n})^2\over n }
   +{(k_BT)^2\over 2\tilde{\epsilon}_1}n_0({\bbox \nabla}_\perp\theta)^2 \\ \nonumber
& & + \int_{{\bf r'}_\perp} V_\perp({\bf r}_\perp-{\bf r'}_\perp)\delta\hat{n}({\bf r}_\perp,z)\delta\hat{n}({\bf r'}_\perp,z)\Bigg\}.
\end{eqnarray}

To Gaussian order the tilt autocorrelator is given by
\begin{equation}
\label{TNgauss}
T_{ij}^0({\bf q}) =
\frac{n_0k_B T}{\tilde{\epsilon}_1}  \delta_{ij} 
+ \; \Big(\frac{n_0k_B T}{{\tilde{\epsilon}_1}}\Big)^2 
  q_{\perp i}q_{\perp j} \langle \theta ({\bf q}_\perp) \theta (-{\bf q}_\perp) \rangle_{G} \;,
\end{equation}
where $\langle ...\rangle_{G}$ denotes an average over the Gaussian action 
(\ref{linearS}).
The  correlation functions of the fluctuating fields are easily calculated within the Gaussian
approximation,
with the result,
\begin{equation}
\label{elcorV}
\langle \delta \hat{n} (-{\bf q}) \delta \hat{n} ({\bf q}) \rangle_{G} = \frac{n_{0} k_B T
q^2 /\tilde{\epsilon}_1}{q_z^2 + \epsilon_B(q)^2/(k_B T)^2} \; ,
\end{equation}
\begin{equation}
\label{elcorVI}
\langle \theta(-{\bf q}) \delta \hat{n} ({\bf q}) \rangle_{G} = \frac{q_z}{q_z^2 + \epsilon_B(q)^2/(k_B T)^2} \; ,
\end{equation}
\begin{equation}
\label{elcorIV}
\langle \theta(-{\bf q})\theta({\bf q}) \rangle_{G} = \frac{\tilde{\epsilon_1}\epsilon_B(q)^2/(n_{0}
q^2 (k_B T)^2)}{q_z^2 + \epsilon_B(q)^2/(k_B T)^2} \; ,
\end{equation}
where
\begin{equation}
\label{Bogol}
\frac{{\epsilon}_B (q_{\perp})}{k_B T} = \Big[ \;  \frac{n_0 k_B T {q_{\perp}}^2 V_\perp
(q_{\perp})}{\tilde{\epsilon}_1} + (\frac{k_B T {q_{\perp}}^2}{2
\tilde{\epsilon}_1})^2 \; \Big]^{1/2}
\end{equation}
corresponds to  the Bogoliubov spectrum of the two-dimensional boson superfluid.
The quartic term in the Bogoliubov spectrum arises from the $ |{\bbox \nabla}_{\perp}\hat{n}|^2 $ 
``kinetic'' term in the action.
To this Gaussian order of approximation the tilt modulus is 
dispersionless and simply the bare part of the single-vortex
contribution to $c_{44}$,
given by
\begin{equation}
c_{44}^0=c_{44}^{v0}=n_0\tilde{\epsilon}_1,
\end{equation}
as given in Eq. (\ref{tiltv}).
By comparing the correlation functions given in Eqs. 
(\ref{elcorV}-\ref{elcorIV}) to those of the hydrodynamic fields
given in Eqs. (\ref{elcorII}-\ref{elcorI}), 
we see that the results obtained by these two methods 
agree with each other
provided we drop the term of ${\cal O}(q_\perp^4)$
in the Bogoliubov spectrum (which is of higher order in the wave vector
and therefore is consistently neglected in a long wavelength theory)
and make the identifications
$c_{44}^0(q_\perp,q_z)=n_0\tilde{\epsilon}_1$ and $c_{11}^0(q_\perp,q_z)=n_0^2V_\perp(q_\perp)$. 
The quantity that replaces the ``Bogoliubov spectrum''
in hydrodynamics is a characteristic inverse length scale
$\xi_z^{-1}$ that controls the decay of correlations along the
$z$ direction, given by 
\begin{equation}
\label{xicorr}
\bigg[{\epsilon(q_\perp)\over k_BT}\bigg]^{1/2}\rightarrow
\xi_z^{-1}(q_\perp,q_z)=q_\perp\sqrt{c_{11}^0(q_\perp,q_z)\
\over c_{44}^0(q_\perp,q_z)}.
\end{equation}
Notice, however, that, in contrast to the boson spectrum, the correlation 
length $\xi_z$ depends on $q_z$, not just on $q_\perp$. This dependence arises
from the nonlocality of the intervortex interaction in the field direction
and will have important consequences on the renormalization of $c_{44}$.
Finally, we stress that the hydrodynamic tilt field does {\it not}
simply map onto the momentum density of two-dimensional bosons, 
which in turn is related to the boson phase variable by
Eq. (\ref{current}). The boson momentum density is to lowest order purely
longitudinal while the tilt field always has a transverse part.

T\"auber and Nelson evaluated perturbatively the corrections to 
$c_{44}^v$ arising from 
terms beyond Gaussian in the free energy \cite{TN}. These corrections can
be obtained by factorizing  the fourth order correlator on the right hand
side of Eq. (\ref{TNresult}) as a product of Gaussian correlators using Wick's
theorem \cite{footb}. For the long wavelength tilt modulus, these
authors obtained
\begin{eqnarray}
\label{TNtiltq}
{1\over c_{44}^{vR}}= {1\over n_0 \tilde{\epsilon}_1} 
\Big[1 - {n_n^B\over n_0}\Big],
\end{eqnarray}
where 
\begin{equation}
\label{superfld}
n_n^B = \frac{L k_B T}{8\tilde{\epsilon}_1} \int \frac{d^2 {\bf q}_{\perp}}{(2 \pi)^2}
\bigg[\frac{q_{\perp}}{\sinh \frac{L {\epsilon}_B (q_{\perp})}{2 k_B T}}\bigg]^2  
\end{equation}
is the normal-fluid density of the two-dimensional boson liquid.
The long-wavelength tilt modulus can also be written as
\begin{equation}
\label{TNtiltlong}
{\frac{1}{c_{44}^{vR}}} = {n_s^B\over n_0^2\tilde{\epsilon}_1}\;,
\end{equation}
where $n_s^B=n_0-n_n^B$ is the boson superfluid density.
As easily seen from Eq. (\ref{superfld}) and discussed in TN \cite{TN}, 
the normal-fluid density is finite only for 
samples of finite thickness $L$,
corresponding to a nonzero boson temperature. In this case one obtains a
renormalization of the tilt modulus due to finite-size effects.
The sign of this correction is
sensitive to the choice of boundary conditions
(the result for periodic boundary conditions is displayed here).
The normal fluid density  vanishes, however,
for $\ L \rightarrow \infty$. The local boson model therefore predicts
that the tilt modulus of an infinitely thick, clean superconductor
is unrenormalized and equals its bare value $n_0\tilde{\epsilon}_1$.
In other words, the flux-line liquid is always entangled in the
thermodynamic limit.

%
%\end{multicols}
%\begin{eqnarray}
%\label{TNnormalf}
%n_n^B(q_\perp,q_z)=
%   {k_BT\over n_0\tilde{\epsilon}_1 LA}\sum_{{\bf q}'_\perp, q'_z}\bigg\{ & &
%     {{q'}_\perp^2\over {q'}_{z}^2+[\epsilon_B({q'}_\perp)/k_BT]^2} 
%    - {({\bf q}_\perp -{\bf q}'_\perp)^2
%     \over (q_z-q'_z)^2+[\epsilon_B(|{\bf q}_\perp-{\bf q'}_\perp|)/k_BT]^2}
% \\ \nonumber
%& &  {({\bf\hat{q}}_\perp\cdot{\bf\hat{q}'}_\perp)^2({\bf q}_\perp-{\bf q'}_\perp)^2
%     [\epsilon_B({q'}_\perp)/k_BT]^{2} 
%        -\big[1-({\bf\hat{q}}_\perp\cdot{\bf\hat{q}'}_\perp)^2\big]{q'}^2_\perp
%              {q'}_z({q'}_z-q_z)\over 
%     \big[{ q'}_{z}^2+[\epsilon_B({q'}_\perp)/k_BT]^{2}\big]
%      \big[ (q_z-q'_z)^2+[\epsilon_B(|{\bf q}_\perp-{\bf q'}_\perp|)/k_BT]^{2}\big]}
%  \bigg\},
%\end{eqnarray}
%
%\begin{multicols}{2}

\section{Non-Gaussian hydrodynamics and disentangled flux liquids}

Our goal in the remainder of this paper is to construct a {\it non-Gaussian}
fully {\it nonlocal} hydrodynamic theory and use it to evaluate the
renormalization of the tilt modulus. As a first step in this direction,
in this section we derive  a non-Gaussian hydrodynamic free energy
from the {\it local} boson action given in Eq. (\ref{actionntheta}). Of course
such a hydrodynamic theory neglects interactions that are nonlocal in
$z$ and will mainly be used as a guide for constructing a 
more general non-Gaussian nonlocal hydrodynamics in the next section.
The non-Gaussian terms in the free energy renormalize the tilt
modulus. When these corrections are evaluated perturbatively, 
the resulting $c^R_{44}$ is identical to that obtained by by T\"auber
and Nelson using the boson formalism \cite{TN}. The main goal of this section
is to emphasize the relationship between the boson formalism and
hydrodynamics and to stress that equivalent results can be obtained by either
method.

To derive the hydrodynamic free energy from the boson action,
we employ  the method used by Kamien and collaborators \cite{kamien} for the formally analog problem
of directed polymers in a nematic solvent. We begin by 
 eliminating
the term ${(k_BT)^2\over 2\tilde{\epsilon}_1}\hat{n}({\bbox \nabla}_\perp\theta)^2$ 
in Eq. (\ref{nonlinearS})
in favor of a new vector
field ${\bf P}$, via a Hubbard-Stratonovich transformation, with the result
\begin{equation} 
{\cal Z}_{\rm gr}({\bf H})=\int{\cal D}\hat{\bf P}{\cal D}\hat{n}{\cal D}\theta e^{-{\cal
S}'[\hat{\bf P},\hat{n},\theta;{\bf h}]/{k_B T}},
\end{equation}
where
\end{multicols}
\begin{eqnarray}
{\cal S}'[\hat{\bf P},\hat{n},\theta;{\bf h}]= & &\int_{\bf r} \bigg\{
ik_BT\hat{n}\partial_z\theta 
  +{(k_BT)^2\over 8\tilde{\epsilon}_1}{({\bbox \nabla}_\perp \hat{n})^2\over \hat{n}}
  +{(k_BT)\over \tilde{\epsilon}_1}i\hat{n}{\bbox \nabla}_\perp\theta\cdot\big[k_BT\hat{\bf P}-{\bf h}\big] \\ \nonumber
& &      +{\hat{n}\over 2\tilde{\epsilon}_1}\big[(k_BT\hat{\bf P})^2-h^2\big]
+ \int_{{\bf r'}_\perp} V_\perp({\bf r}_\perp-{\bf r'}_\perp)\hat{n}({\bf r}_\perp,z)\hat{n}({\bf r'}_\perp,z)\bigg\}.
\label{nonlinearSp}
\end{eqnarray}
If we integrate over $\hat{\bf P}$ in Eq. (\ref{nonlinearSp}), we return to the 
original nonlinear action. Instead we integrate over $\theta$ which only appears linearly
in the new action. This integration results in a $\delta$-functional,
yielding
\begin{equation}
\tilde{\cal Z}_{\rm gr}({\bf H})=\int {\cal D}\hat{n}{\cal D}\hat{\bf P} 
\exp\Big[{2 k_BTn_0^2\over\tilde{\epsilon}_1}\int_{\bf r}\ln(\hat{n}({\bf r})/n_0)\Big]
e^{-\tilde{\cal S}_H[\hat{n},\hat{\bf P};{\bf h}]/{k_B T}}
 \delta\big(\partial_z \hat{n}+
{\bbox \nabla}_\perp\cdot{n\over\tilde{\epsilon}_1}(k_BT\hat{\bf P}+{\bf h})\big),
\label{hydroZ}
\end{equation}
with
\begin{equation}
\tilde{\cal S}_H[\hat{n},\hat{\bf P};{\bf h}]=  {1\over 2}\int_{\bf r}\Big\{{(k_BT)^2\over\tilde{\epsilon}_1}\hat{n}\hat{\bf P}^2+
{(k_BT)^2\over 4\tilde{\epsilon}_1}{({\bbox \nabla}_\perp \hat{n})^2\over \hat{n}}
  -{\hat{n}\over2\tilde{\epsilon}_1}h^2
 + \int_{{\bf r'}_\perp} V_\perp({\bf r}_\perp-{\bf r'}_\perp)\hat{n}({\bf r}_\perp,z)\hat{n}({\bf r'}_\perp,z)\Big\}.
\label{hydroS}
\end{equation}
\begin{multicols}{2}
In obtaining Eq. (\ref{hydroS}) we have discretized the nonlinear action 
(\ref{nonlinearS}) in real space, according to
\begin{equation}
\label{discretize}
\int_{\bf r} f({\bf r}) \rightarrow v_0\sum_i f_i,
\end{equation}
with $v_0$ an elementary volume, $v_0=\tilde{\epsilon_1}/(2 k_{B} T
n_{0}^2)$. This is the volume of a box with base area equal to $1/n_{0}$
and height equal to the single-vortex entanglement length,
\begin{equation}
\label{entlength}
l_z={\tilde{\epsilon_1}\over2 k_{B} T n_{0}}.
\end{equation}
The term containing the logarithm of the fluctuating density
arises from the Jacobian of the functional integration over the full
nonlinear action.
It represents the nonlinear ``ideal gas''
part of the flux liquid free energy.

Statistical averages have to be performed by integrating over the fields
$\hat{n}({\bf r})$ and $\hat{\bf P}({\bf r})$ with the constraint provided
by the $\delta$-functional in Eq. (\ref{hydroZ}).
Comparison of Eq. (\ref{hydroZ}) 
to the hydrodynamic free energy (\ref{linhydro}) of a flux-line liquid 
with the constraint (\ref{constraint}) suggests a
physical interpretation for the auxiliary vector field $\hat{\bf P}$.
The quantity $\hat{n}(k_BT\hat{\bf P}+{\bf h})\tilde{\epsilon}_1$ takes the place of the
hydrodynamic 
tilt field $\hat{\bf t}^H$ introduced in the previous section. 
The difference between the vector field $\hat{\bf P}$ and the tilt 
field can be understood by 
noting that,
as pointed out by Nelson and Le Doussal \cite{drn_ledoussal},
the canonically conjugate momentum of the fictitious particle that
corresponds to the  $n$-th flux-line
is ${\bf p}_n=i\big(\tilde{\epsilon}_1{d{\bf r}_n\over dz}+{\bf h}\big)$.
The vector field $\hat{\bf P}$ can then be interpreted as a 
sort of ``velocity'' field, while the
tilt field $\hat{\bf t}^H$ represents the canonically conjugate momentum density.
The two differ in the presence of an applied transverse field
${\bf h}$ that contributes to the single-vortex ``canonical momentum''.

The relationship between the effective action $\tilde{\cal S}_H$ and
the hydrodynamic free energy of a tilted flux-line liquid is made
more transparent by performing and additional change of variable
that replaces the field $\hat{\bf P}$ by a tilt field defined as
\begin{equation}
\label{chvarII}
\hat{\bf t}({\bf r})=\frac{\hat{n}({\bf r})}{\tilde{\epsilon}_1}\big[k_B T  \hat{\bf P}({\bf r}) + {\bf h}({\bf r})\big].
\end{equation}
The Jacobian of this transformation  cancels the Jacobian of the 
Hubbard-Stratonovich transformation used earlier and we obtain,
\end{multicols}
\begin{eqnarray}
\label{tnZ}
{\cal Z}_{gr}({\bf H}) = \int {\cal D} \hat{n} {\cal D}\hat{\bf t}
e^{-{\cal S}_H[\hat{n},\hat{\bf t};{\bf h}]/k_BT}\delta\big(\partial_z \hat{n}+
{\bbox \nabla}_\perp\cdot\hat{\bf t}\big) \;,
\end{eqnarray}
with
\begin{equation}
\label{hydaction}
{\cal S}_H[\hat{n},\hat{\bf t};{\bf h}]={1\over 2k_BT}
\int_{\bf r}\Big[{\tilde{\epsilon}_1}\frac{\hat{\bf t}^2}{\hat{n}} +
{(k_BT)^2\over 4\tilde{\epsilon}_1}{({\bbox \nabla}_\perp \hat{n})^2\over \hat{n}} \nonumber\\
-{\bf h}\cdot{\bf t} 
 + \int_{{\bf r'}_\perp} V_\perp({\bf r}_\perp-{\bf r'}_\perp)\hat{n}\big({\bf r}_\perp,z)\hat{n}({\bf r'}_\perp,z)\Big].
\end{equation}
\begin{multicols}{2}
The effective action  of a tilted flux-line liquid given in
Equation (\ref{hydaction}) becomes formally identical to the corresponding
nonlinear hydrodynamic free energy, provided we make the identifications,
\begin{eqnarray}
\label{boshyd}
& & \hat{n}({\bf r})\leftrightarrow \hat{n}^H({\bf r}),\\ \nonumber
& & \hat{\bf t}({\bf r})\leftrightarrow \hat{\bf t}^H({\bf r}),\\ \nonumber
& & n_0 {\tilde{\epsilon}_1} \leftrightarrow c_{44}^0({\bf q}),\\ \nonumber
& & {n_0}^2 V_\perp(q_\perp)\leftrightarrow c_{11}^{0}({\bf q}).
\end{eqnarray}
The corresponding  hydrodynamic free energy is 
nonlinear, but local in $z$, and it given by
\begin{equation}
\label{localhyd}
F^{\ell}[\hat{n}^H,\hat{\bf t}^H;{\bf h}]=k_BT{\cal S}_H[\hat{n},\hat{\bf t};{\bf h}].
\end{equation}
The subscript ``$\ell$'' indicates that only local interaction among the
vortices has been retained in this hydrodynamic free energy. The free energy $F^l$ contains the term quadratic in the density gradient that is
neglected in conventional hydrodynamics.
We will retain this term here to make our
comparison with the
results of the boson theory more transparent. Also this term will
be needed below to provide a large wave vector cutoff to the integrals
determining the renormalized tilt modulus.

The long wavelength part of the tilt-tilt autocorrelator can now be evaluated using the definition,
Eq. (\ref{tiltcorrder}).
The non-Gaussian terms in the local hydrodynamic free energy (\ref{localhyd})
are separated out by writing
\begin{equation}
F^{\ell}=F_G^{\ell}+\delta F^{\ell},
\end{equation}
where $F_G^{\ell}$ is given by Eq. (\ref{linhydro}), but with the values specified
in Eqs. (\ref{boshyd}) for the elastic constants, and
\begin{equation}
\label{nonGhydro}
\delta F^{\ell}=-{\tilde{\epsilon}_1\over 2n_0}\int_{\bf r}\hat{\bf t}^2{\delta \hat{n}\over
 \hat{n}}.
\end{equation}
The tilt autocorrelator is then evaluated perturbatively in 
the non-Gaussian part $\delta F^{\ell}$ of the free energy. The
perturbation expansion is outlined in Appendix B. 
To leading order,
we obtain
\begin{equation}
T_{ij}({\bf q})=T_{ij}^0({\bf q})+\delta T_{ij}({\bf q}),
\end{equation}
where $T_{ij}^0({\bf q})$ is the bare part of the correlator, given by Eq. 
(\ref{elcorI}-\ref{longittilt}).
The hydrodynamic limit of the correction $\delta T_{ij}({\bf q})$ is given by
\end{multicols}
\begin{equation}
\lim_{q_z \rightarrow 0}\delta T_{ij}(0, q_z)={\frac{n_{0} k_B
T}{\tilde{\epsilon_1}}}\delta_{ij}-{\frac{(k_B T)^2}{{\tilde{\epsilon_1}}^2
LA}}\sum_{{\bf q}'_{\perp},q'_z} q'_i q'_j {\frac{({\epsilon}_B ({q'}_{\perp})/(k_B T))^2-{q'}_z^2}{[({\epsilon}_B ({q'}_{\perp})/(k_B T))^2+{q'}_z^2]^2}}.
\end{equation}
\begin{multicols}{2}
This result is identical to that obtained 
obtained by T\"auber and Nelson via the boson formalism.
In particular, the long wavelength tilt modulus 
defined according to Eq. (\ref{longtilt}) is found to be 
given by Eq. (\ref{TNtiltq}), with
\end{multicols}
\begin{equation}
n_n^B={\frac{n_0k_B T}{2LA}}
\sum_{{\bf q}_{\perp},q_z}q_{\perp}^2 {\frac{({\epsilon}_B (q_{\perp})/(k_B T))^2-{q}_z^2}{[({\epsilon}_B (q_{\perp})/(k_B T))^2+{q}_z^2]^2}},
\end{equation}
which becomes identical to Eq. (\ref{superfld}) in the thermodynamic limit of large sample size. 
\begin{multicols}{2}

\section{Tilt modulus from  nonlocal, non-Gaussian  hydrodynamics}

As discussed in the Introduction, neglecting the interaction 
among vortex segments at different ``heights'' $z$ has severe
effects on
the flux liquid tilt modulus, namely it completely neglects
its  compressional part, which is the largest contribution over a 
wide part of the $(H,T)$ phase diagram. Hence our desire to 
develop a simple formalism for the calculation of the tilt modulus of a flux-line liquid that incorporates such nonlocalities.

A generalization of the boson mapping that incorporates the $z$-nonlocality of the vortex interaction was proposed some time ago by
Feigel'man and collaborators \cite{feigel}. The $z$-nonlocality yields a retarded
interaction among the bosons that can be handled by the introduction
of a Chern-Simons gauge field. In the limit of infinite penetration depth,
$\lambda_\perp$, considered by these authors, the flux-line array then
maps onto a {\it charged} superfluid. These authors argued that the charged
boson system possesses a normal-fluid phase at zero temperature,
corresponding to a thermodynamically distinct
disentangled flux liquid phase, with infinite tilt modulus
and longitudinal superconductivity.

Nonlocality is incorporated in a natural way in hydrodynamics.
A {\it nonlinear} hydrodynamic free energy that incorporates
all nonlocalities of the intervortex interaction can be obtained phenomenologically 
by coarse-graining of the microscopic energy of the vortex liquid,
following the methods described in Ref. \onlinecite{mcment}.
Care must be taken in handling the self-interaction between segments of
the same flux-line at different $z$ heights, which is responsible for
the non-Gaussian terms in the hydrodynamic free energy.
Such non-Gaussian terms are neglected in the linearized theory,
but as seen in the previous section they control the renormalization of the tilt modulus.
The nonlinear hydrodynamic free energy obtained by such a procedure is given by
\end{multicols}
\begin{equation} 
\label{nlF}
F=\frac{1}{2n_0^2}\int_{\bf r} \int_{{\bf r}'} \Bigg\{ \Big[
\frac{n_0^2\tilde{\epsilon}_1}{\hat{n}^H({\bf r})} \delta ({\bf r} - {\bf r}') +
K_c ({\bf r} - {\bf r}') \Big] 
\hat{\bf t}^H({\bf r})\hat{\bf t}^H({\bf r}') + B({\bf r} - {\bf r}') \delta \hat{n}^H ({\bf
r})\delta \hat{n}^H ({\bf r}')\Bigg\} ,
\end{equation}
\begin{multicols}{2}
\noindent where $B({\bf r})$ is the real space compressional modulus
and $K_{c}({\bf r})$  is the interaction part of the real space tilt modulus. The first term in Eq. (\ref{nlF}) arises from the self-energy part
of the interaction and it represents a sort of nonlinear 
``kinetic'' contribution to the total energy of the flux-line array.
To make contact with conventional notation, it is convenient
to rewrite the interaction part of the free energy in wave-vector
space,
\begin{eqnarray}
\label{nonlinhydF}
F = & & {1\over 2 }\int_{\bf r} \tilde{\epsilon}_1
\frac{[\hat{\bf t}^H({\bf r})]^2}{\hat{n}^H({\bf r})} \\ \nonumber 
& & + {1\over 2 n_0^2\Omega} \sum_{\bf q} \Big\{ c_{44}^{c0}({\bf q})
|\hat{\bf t}^H({\bf q})|^2 
+ c_{11}^0({\bf q}) | \delta \hat{n}^H ({\bf q})|^2  \Big \} \,
\end{eqnarray}
where the bare compressional modulus, $c_{11}^0({\bf q})$, and
the interaction part of the bare tilt modulus, $c_{44}^{c0}({\bf q})$,
are given in Eqs. (\ref{barecomp}) and (\ref{tiltc}), respectively. 

The non-Gaussian hydrodynamic free energy can also
be derived from the action of two-dimensional
bosons with retarded interaction written down by Feigel'man and collaborators
by successively eliminating nonhydrodynamic fields in favor of hydrodynamic
fields via formal manipulations analogous to those described in the 
previous section. This derivation is outlined in Appendix A.
The resulting free energy differs from the phenomenological one given
in Eq. (\ref{nonlinhydF}) only in that it contains an additional term
proportional to density gradients (see Appendix A). This term is usually neglected
in hydrodynamics because it is of higher order in the gradients. We will,
however, retain it here as it provides an intrinsic large-wave-vector cutoff
to the integrals determining the renormalized tilt modulus. 
It can be incorporated in the free energy of Eq. (\ref{nonlinhydF})
by the replacement 
\begin{equation}
c_{11}^0({\bf q})\rightarrow c_{11}^0({\bf q})+(k_BT)^2n_0q_\perp^2/(4\tilde{\epsilon}_1).
\end{equation}
It is convenient for the following to separate out the non-Gaussian
part of the hydrodynamic free energy of Eq. (\ref{nonlinhydF})
by letting 
\begin{equation}
\label{splitF}
F=F_G+\delta F,
\end{equation}
where $F_G$ is given by Eq. (\ref{linhydro}), and
\begin{equation}
\label{deltaF}
\delta F=-{1\over 2 }\int_{\bf r} {\tilde{\epsilon}_1 [\hat{\bf t}^H({\bf r})]^2
\over n_0}{\delta \hat{n}^H({\bf r})\over \hat{n}^H({\bf r})}.
\end{equation}
The tilt autocorrelator can be evaluated by treating 
the non-Gaussian part of the free energy (\ref{deltaF}) perturbatively. 
Some details are
given in Appendix B. The dimensionless parameter that
controls the expansion in $\delta F/k_BT$
is proportional to $(\tilde{\epsilon}_1 /2 k_B T
\sqrt{n_0})^2=(l_z/a_0)^2$, with $l_z$ the entanglement
length given in Eq. (\ref{entlength}).
Small values of $l_z/a_0$ correspond to
an entangled flux-line liquid. The ``kinetic'' nonlinearities that
are incorporated perturbatively 
stiffen the tilt modulus of the line liquid, making it therefore less
entangled. 

The nonlinearities embodied in $\delta F$ yield corrections to all
the correlation functions. Here, we only display the result for the 
transverse part of the tilt-tilt correlator, 
that determines the wave vector-dependent tilt 
modulus. Using Eq. (2.30), the wave vector-dependent  tilt modulus is
given by
\begin{equation}
\label{c44ans}
{1\over c_{44}^R(q_\perp, q_z)}={1\over c_{44}^0(q_\perp, q_z)}
\bigg[1-{n_0\tilde{\epsilon}_1\over c_{44}^0(q_\perp, q_z)}
{n_n(q_\perp, q_z)\over n_0}\bigg],
\end{equation}
with
\end{multicols}
\begin{eqnarray}
\label{nfans}
n_n(q_\perp,q_z)& =& {k_BT\over LA}\sum_{{\bf q}'_\perp, q'_z}\bigg\{
     {{q'}_\perp^2\over c_{44}^0({\bf q'})}
     {1\over {q'}_{z}^2+[\xi_z({\bf q'})]^{-2}} 
    - {n_0\tilde{\epsilon}_1({\bf q}_\perp -{\bf q}'_\perp)^2
           \over c_{44}^0({\bf q'})
                 c_{44}^0({\bf q}-{\bf q}')}
     {1\over (q_z-q'_z)^2+[\xi_z({\bf q}-{\bf q'})]^{-2}}\bigg\} \\ \nonumber
& &  +{n_0\tilde{\epsilon}_1k_BT\over LA}\sum_{{\bf q}'_\perp, q'_z}
     {({\bf\hat{q}}_\perp\cdot{\bf\hat{q}'}_\perp)^2({\bf q}_\perp-{\bf q'}_\perp)^2
     [\xi_z({\bf q'})]^{-2} 
        -\big[1-({\bf\hat{q}}_\perp\cdot{\bf\hat{q}'}_\perp)^2\big]{q'}^2_\perp
              {q'}_z({q'}_z-q_z)\over c_{44}^0({\bf q'})
                    c_{44}^0({\bf q}-{\bf q'})
     \big[{ q'}_{z}^2+[\xi_z({\bf q'})]^{-2}\big]
      \big[ (q_z-q'_z)^2+[\xi_z({\bf q}-{\bf q'})]^{-2}\big]},
\end{eqnarray}
and 
\begin{equation}
[\xi_z({\bf q})]^{-2}={q_\perp^2\over c_{44}^0({\bf q})}
       \Big[c_{11}^0({\bf q})
     +{(k_BT)^2n_0q_\perp^2\over 4\tilde{\epsilon}_1}\Big].
\end{equation}
The length scale $\xi_z({\bf q})$ differs from the one defined in 
Eq. (\ref{xicorr}) in that it contains an additional term arising from the
coupling to the density gradient contained in our  free
energy and usually neglected in hydrodynamics. For simplicity,
we use, however, the same notation as in Eq. (\ref{xicorr}).

The long-wavelength tilt modulus is determined by 
$n_n=\lim_{q_z\rightarrow 0}\lim_{q_\perp\rightarrow 0}n_n(q_\perp,q_z)$, given by
\begin{equation}
\label{nf}
n_n ={k_BT\over LA}\sum_{{\bf q}_\perp, q_z}
     {q_\perp^2\over c_{44}^0({\bf q})}
     \bigg[1-{n_0\tilde{\epsilon}_1\over c_{44}^0({\bf q})}\bigg]
     {1\over q^2_z+[\xi_z({\bf q})]^{-2}} 
    +{n_0\tilde{\epsilon}_1k_BT\over 2LA}\sum_{{\bf q}_\perp, q_z}
     {q_\perp^2\over [c_{44}^0({\bf q})]^2}
     {[\xi_z({\bf q})]^{-2}-q_z^2\over\big[ q^2_z+[\xi_z({\bf q})]^{-2}
     \big]^2}.
\end{equation}
\begin{multicols}{2}

Equations (\ref{c44ans}-\ref{nf}) are the central result
of this paper. 
If the $z$-nonlocality of the intervortex interaction is neglected in 
Eq. (\ref{nfans}) by replacing the
elastic constants on the right-hand side 
with the corresponding values used in the
the local boson formalism,
according to Eq. (\ref{boshyd}), then Eq. (\ref{nfans})
becomes identical to the result obtained by TN.
In particular, the first term on the right hand side of Eq. (\ref{nf})
is absent in the local boson model of TN, where $c_{44}^0=n_0\tilde{\epsilon}_1$.
The long-wavelength normal fluid density is then
given by Eq. (\ref{superfld}) and  vanishes for $L\rightarrow\infty$.

The normal fluid density given in Eq. (\ref{nf}) can be
evaluated explicitly for the case of an isotropic
superconductor ($p=1$) in the limit of infinite thickness ($L\rightarrow\infty$).
After inserting in  Eq. (\ref{nf}) the expression for the nonlocal 
bare elastic constants 
given in Eqs. (\ref{barecomp}) and (\ref{tilt}-\ref{tiltc}), the $q_z$ integral 
in Eq. (\ref{nf}) can be evaluated.
The resulting normal-fluid fraction depends on the three length scales that characterize the system.
These are the
average intervortex spacing, $a_0=1/\sqrt{n_0}$, the the $ab$ plane
London penetration depth, 
$\tilde{\lambda}_{\perp}$, and the single-vortex entanglement length,
$l_{z}$. We have introduced two 
dimensionless parameters,
\begin{equation}
\label{u}
u={2l_z\over\sqrt{\pi}a_0}={2\tilde{\epsilon}_1\over{k_{B}T\sqrt{4\pi n_{0}}}},
\end{equation}
and a dimensionless volume fraction of vortex lines,
\begin{equation}
\label{v*}
v^{*}={1\over{4\pi n_0{\tilde\lambda}_\perp^2}},
\end{equation}
The renormalized long-wavelength tilt modulus is written in terms of our dimensionless
parameters as
\begin{equation}
\label{rtilt0}
{1\over c_{44}^R}={1\over c_{44}^0}\bigg[1-{v^*\over 1+v^*}{n_n\over n_0}\bigg]
\end{equation}
and the normal fluid fraction is given by
\begin{equation}
\label{infthickres}
{n_n\over{n_0}}={1\over{2u}}\int_0^{\infty}dx\{K(x|u,v^*)+L(x|u,v^*)\}\;,
\end{equation}
where
\end{multicols}
\begin{eqnarray}
\label{K}
K(x|u,v^*)={\frac{x^2[1+(x+v^{*})(1+x/u^2)]+2z_{1}z_{2}x(x+v^{*})}{\sqrt{x+v^{*}}z_{1}z_{2}(z_{1}+z_{2})[\sqrt{1+x+v^{*}}(x+z_{1}z_{2})+z_{1}z_{2}(z_{1}+z_{2})]}}\;,
\end{eqnarray}
\begin{eqnarray}
\label{L}
L(x|u,v^*)=v^*{\frac{x(z_1^2+z_2^2)}{z_{1}z_{2}(z_{1}+z_{2})(z_1^2-z_2^2)}}\;,
\end{eqnarray}
with
\begin{eqnarray}
\label{z1}
z_{1,2}={1\over{\sqrt{2}}}\{1+x+(x/u)^2+v^{*}\pm[(1+(x/u)^2-x-v^{*})^2+4v^{*}]^{1/2}\}^{1/2}\;.
\end{eqnarray}
\begin{multicols}{2}
These integrals have been evaluated numerically. The resulting normal
fluid fraction is shown in Fig. 1 as a function of $u$ for several values 
of the volume fraction $v^*$. We note that the dependence on $v^*$ is rather weak,
particularly for small values of $u$. 

%*********Figure 1**********
%
\begin{figure}[bth]
{\centering
\setlength{\unitlength}{1mm}
\begin{picture}(150,70)(0,0)
\put(-3,-25){\begin{picture}(150,70)(0,0) 
\includegraphics{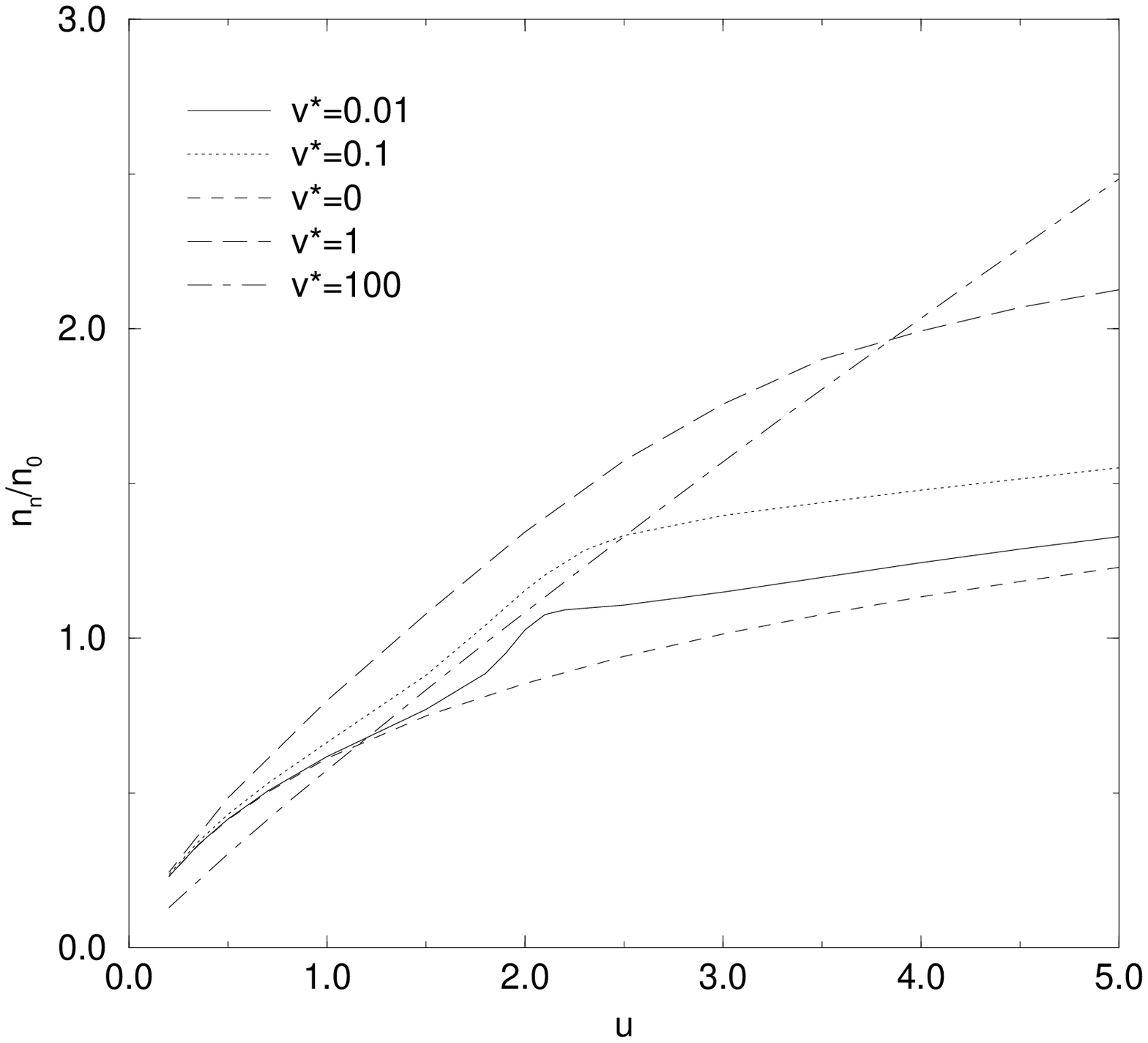}
\end{picture}}
\end{picture}}
Fig.1.{The normal-fluid fraction given by Eq. (7.13) as a function
of $u$ for five
different values of $v^*$. Notice the weak dependence of $n_n/n_0$ on $v^*$
for small values of $u$.}
\end{figure}

For  $v^{*}=0$ (which can be interpreted as either the high density limit or
the infinite $\lambda_\perp$ case
treated by Feigel'man and collaborators \cite{feigel}), the normal-fluid density given
in Eq. (\ref{infthickres}) reduces {--} up to an overall factor of 2 {--} 
to the result obtained by Feigel'man et al. 
\cite{feigel,factoroftwo}. 
Our Eq. (\ref{nf}) generalizes the result obtained by Feigel'man 
and coworkers to the case of finite penetration depth. 

We stress that our calculation is perturbative and we have only evaluated 
the leading correction in perturbation theory. As discussed above,
the small parameter in the perturbation theory is proportional to
$u^2\sim (l_z/a_0)^2$. In other words, the unperturbed state is an
entangled flux liquid, with a very small value of the $z$-axis coherence 
length $l_z$, and interactions stiffen the vortices, enhancing 
the tilt modulus. 
We can estimate the values of magnetic field and temperature  where our 
perturbation theory breaks down as determined by the root of the equation
\begin{equation}
\label{root}
{v^*\over 1+v^*}{n_n\over n_0}=1. 
\end{equation}
The solution $u_0(v^*)$ of Eq. (\ref{root})
defines a line $B_{D0}(T)$ in the $(H,T)$ phase diagram
that can be interpreted as an estimate of the phase boundary between
entangled and disentangled liquid regions. For $B>B_{D0}(T)$ the liquid is 
entangled, while for $B<B_{D0}(T)$ the perturbation theory breaks down,
signaling the appearance of a disentangled flux-line liquid. Of course,
in order to interpret the region $B<B_{D0}(T)$ as a disentangled flux liquid
the $B_{D0}(T)$ line must lie in the molten region of the $(H,T)$
phase diagram.
At high density,
$v^*<<1$ and Eq. (\ref{root}) can be approximated as  
$n_n/n_0\sim 1/v^*>>1$. It is clear from Fig. 1
that the roots of this equation occur at large values of $u$, where $n_n/n_0\sim(1/2)ln(u)$.
We then estimate that 
our perturbation theory breaks down for $u_0(v^*)\sim\exp(2/v^*)$.
Converting to field and temperature, this corresponds to
$B_{D0}(T)\sim (H_{c1}/2\ln\kappa)\ln(H_{c1}\phi_0/\pi k_BT 4 p^2\sqrt{\ln\kappa})$,
with $H_{c1}=\phi_0/4\pi\tilde{\lambda}^2_\perp\ln\kappa$. 
Below this line, $c_{44}$ is strongly renormalized upward by interactions
and a  large disentangled flux-line liquid fraction may appear.
Conversely, at low density, $v^*>>1$ and  Eq. (\ref{root}) becomes $n_n/n_0\sim1$.
The solution of this equation depends weakly on $v^*$, as seen from Fig. 1,
and is approximately $u_0\sim 2$, corresponding to 
$B_{D0}(T)\sim (\phi_0/4\pi)( \tilde{\epsilon}_1 /k_B T)^2$. 
This result coincides with the estimate obtained by Feigel'man et al \cite{feigel},
but it applies in a different field regime.
The solution $u_0(v^*)$ of Eq. (\ref{root}) for general values of $v^*$ has been 
obtained numerically and is shown in Fig. 2 as a solid line.
For small $v^*$ (high vortex-line density) Eq. (\ref{root}) predicts 
that the perturbation theory breaks down at very large values of $u$,
in a region that is well beyond its range of applicability.

%*********Figure 2**********
%
\begin{figure}[bth]
{\centering
\setlength{\unitlength}{1mm}
\begin{picture}(150,70)(0,0)
\put(-3,-25){\begin{picture}(150,70)(0,0) 
\includegraphics{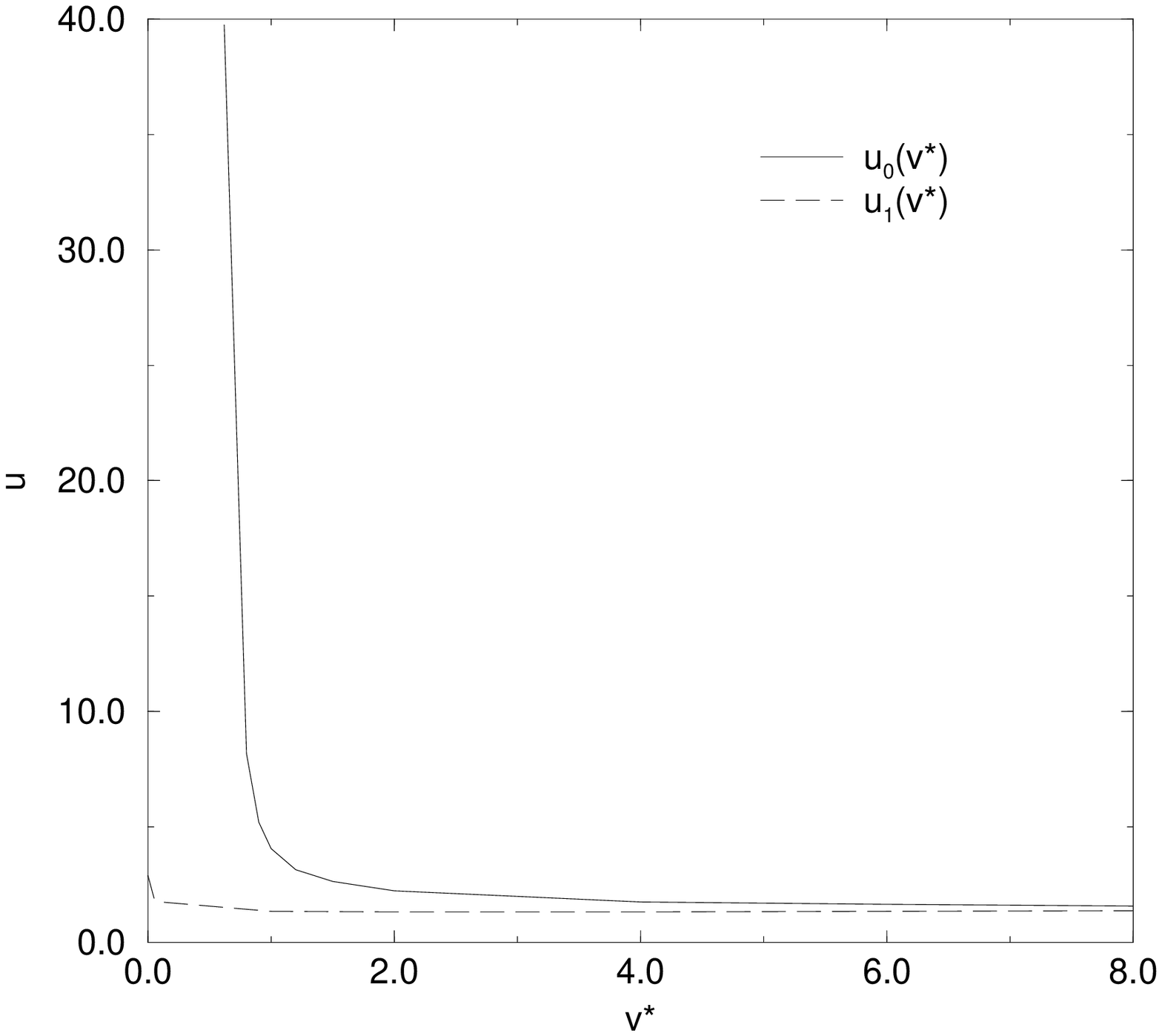}
\end{picture}}
\end{picture}}
Fig.2.{The solid line is the numerical solution of Eq. (7.17). 
It defines the line $u_0(v^*)$ in the $(u,v^*)$
parameter space where the
perturbation expansion of the tilt autocorrelator breaks down. The dashed
line is $u_1(v^*)$, where $n_n/n_0=1$.}
\end{figure}

We now wish to compare our perturbative result to the nonperturbative
expression for $c_{44}$ proposed by Larkin and Vinokur and given in
Eq. (\ref{LVresult}). As discussed in the Introduction, if the Larkin-Vinokur formula is expanded 
for small values of the normal fluid fraction $n_n/n_0$, the leading 
term has the form given in Eq. (\ref{LVapprox}),
which is identical to the long wavelength limit of
our result (\ref{c44ans}), provided we identify $n_n$ in Eq. (\ref{LVapprox})
with our perturbative expression for the normal fluid density
given in Eq. (\ref{nf}).
It is then tempting to conjecture that a nonperturbative generalization of
our calculation may indeed yield the expression (\ref{LVresult}) proposed by Larkin and Vinokur
for the renormalized long wavelength tilt modulus,  but with 
a normal fluid fraction given by Eq. (\ref{nf}), corresponding to
\begin{equation}
\label{conjecture}
{1\over c_{44}^R}={1\over c_{44}^c+{n_0\tilde{\epsilon}_1\over 1-n_n/n_0}},
\end{equation}
with $n_n$ given by Eq. (\ref{nf}). We stress that Eq. (\ref{conjecture}), which is
simply
a rewriting of the Larkin-Vinokur result, 
is purely a conjecture in the context of our work. It is, however, interesting to explore its consequences. 
According to Eq. (\ref{conjecture}), the condition for the vanishing
of $1/c_{44}^R$, corresponding to the onset of a macroscopic
disentangled fluid fraction, would read
\begin{equation}
\label{Feigdisent}
{\frac{n_n}{n_0}}=1\;.
\end{equation}
The numerical solution of this equation, denoted by $u_1(v^*)$,
is shown in Fig. 2 as a dashed line. We note that the line $u_0(v^*)$, where
the perturbation theory breaks down, and the line $u_1(v^*)$, where
the conjectured nonperturbative form of $1/c_{44}^R$ vanishes, coincide
at large $v^*$, but diverge at small $v^*$. In this high density
region it appears that the perturbation theory strongly underestimates
the stiffening of $c_{44}$ from interactions.  The line $u_1(v^*)$ defines
a second ``disentanglement
line'', $B_{D1}(T)$, in the $(H,T)$ phase diagram. 
Assuming $u_1(v^*)\sim 2\sim{\rm constant}$
over the range of $v^*$ values of interest, we estimate 
$B_{D1}(T)\sim (\phi_0/4\pi)( \tilde{\epsilon}_1 /k_B T)^2$. 
Notice that the field $B_{D1}(T)$ (which coincides with $B_{D0}(T)$ 
at low vortex density) is of the order of the melting field $B_{\rm m}(T)$
of the vortex lattice. Using a Lindeman criterion for melting, 
this is found to be 
$B_{\rm m}(T)=(16c_L^4\phi_0p^2/(\ln\kappa)^2(\tilde{\epsilon}_1/k_BT)^2$,
where $c_L$ is the Lindeman parameter \cite{tinkham}.

Before discussing the location of the disentanglement lines
$B_{D0}(T)$ and $B_{D1}(T)$ in the $(H,T)$ phase diagram, we recall that 
the explicit evaluation of the integrals determining the normal fluid density 
has been carried out for isotropic superconductors ($p=1$). To estimate the relevance
of our result to the anisotropic ${\rm CuO}_2$ materials, we have used the above
estimate for the boundary between disentangled and entangled liquid regions and
inserted parameter values typical of these materials. To justify this approximation, we note that
for $p>>1$ the compressional part of the tilt modulus arising from 
the nonlocality of the vortex interaction in the $z$ direction becomes less
important relative to the vortex part. 
As it is precisely this nonlocality that is responsible for a nonvanishing
renormalization of $c_{44}$ in infinitely thick samples,
we expect that the results that we have obtained for the isotropic case will 
provide an upper bound for the size of the renormalization
in anisotropic materials.
A sketch of a phase diagram showing the location of the disentanglement lines
$B_{D0}(T)$ (dashed line) and $B_{D1}(T)$ (dotted line) is shown in Fig. 3. 
It is not drawn to
scale. 

Using parameter values  of ${\rm YBCO}$ and ${\rm BSCCO}$  we have estimated that in both these
materials at high fields $(B>1{\rm Tesla})$ the
$B_{D0}(T)$ boundary defining the breaking down of our perturbation theory
lies well within the flux lattice phase. At low fields 
there is a possibility for a disentangled phase in the reentrant liquid region.
This region is, however,  rather narrow,
particularly in ${\rm YBCO}$ where it is expected
to have a width of the order of $1$ Gauss \cite{footgauss}. For this reason, while we have drwan 
in Fig. 3 the ``horizontal''
part of the $B_{D0}(T)$ curve as passing through this reentrant liquid
phase, it could very well be that this line is located either above (in the lattice) or
below (in the Meissner phase) the sketched position.
The disentanglement line $B_{D1}(T)$ is shown as dotted in Fig. 3 and it is estimated to lie
in the liquid phase. The existence of this line is, however, 
just a conjecture in the context of our work, as our results
are strictly perturbative.
In general we expect the actual disentanglement line to lie between 
our perturbative estimate $B_{D0}(T)$ and the conjectured $B_{D1}(T)$. 
It could therefore lie almost entirely in the solid phase, indicating
that a true thermodynamic disentangled liquid phase does not exist.
This conclusion would appear to agree with the latest results from simulations
\cite{Nguyensudbo,japan,nordborg}.
Further work beyond the naive lowest order perturbation expansion discussed
here is needed, however, to settle this point.

%*********Figure 3**********
%
\begin{figure}[bth]
{\centering
\setlength{\unitlength}{1mm}
\begin{picture}(150,70)(0,0)
\put(-3,-25){\begin{picture}(150,70)(0,0) 
\includegraphics{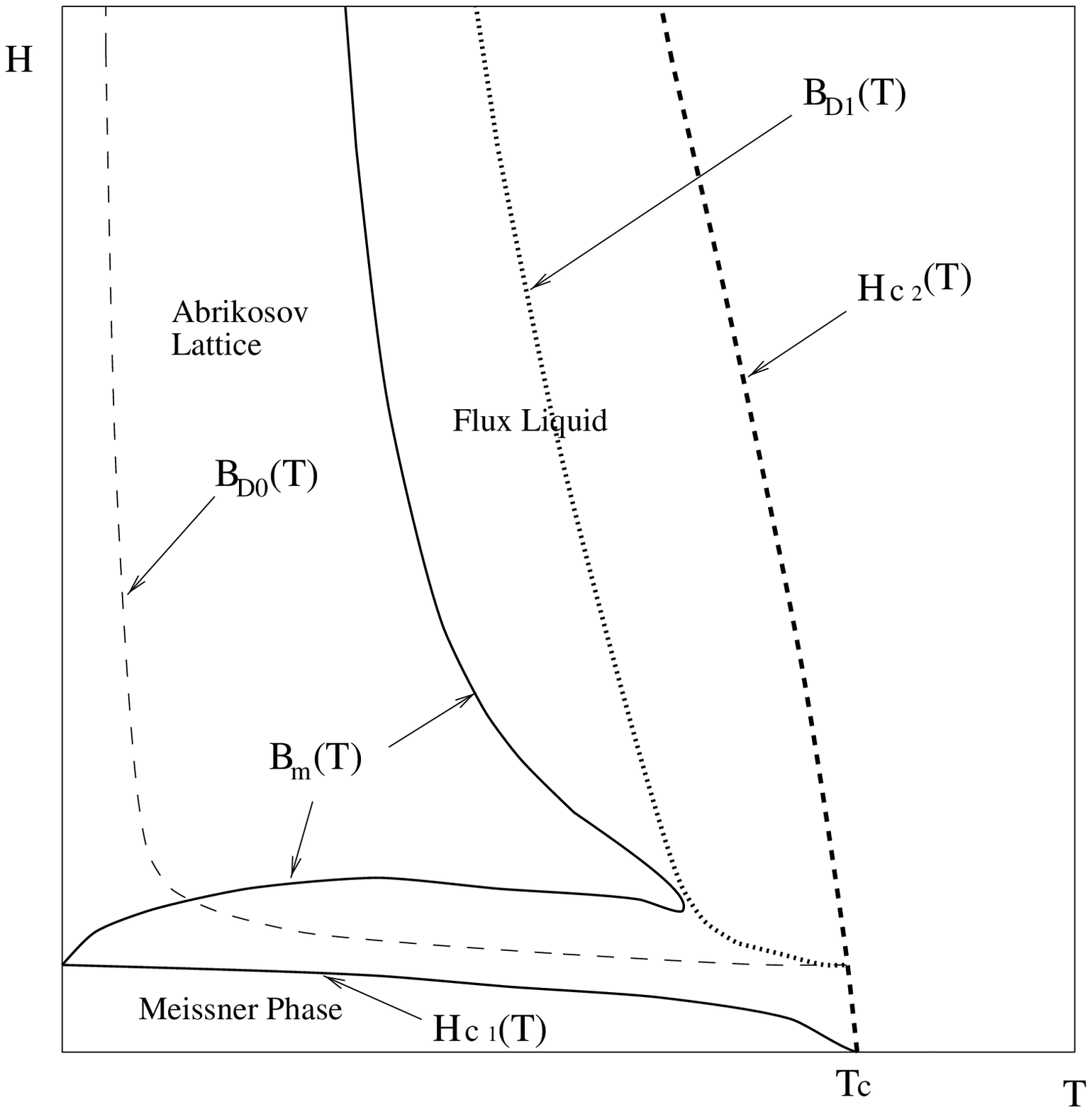}
\end{picture}}
\end{picture}}
Fig.3.{
A sketch (not to scale) of the phase diagram showing
the location of the ``disentanglement'' lines discussed in the text.
The dashed line, $B_{D0}(T)$, marks the breaking down of 
the perturbation expansion for the
inverse tilt modulus; the dotted line, $B_{D1}(T)$, 
corresponds to $n_n/n_0=1$ and signals the divergence of the
conjectured form of $c_{44}^R$, given in Eq. (7.18).
The width of the reentrant liquid phase is in reality much smaller
than shown here  and
the line $B_{D0}(T)$ may or may not pass through it. $B_m(T)$ is
the melting line. $H_{c2}(T)$ marks the onset of a Meissner effect and is
not a sharp phase transition.}
\end{figure}

One important outcome of our work is  that the nonlocality of the intervortex interaction
in the field direction has important qualitative effects on the tilt
modulus. In particular, it always yields a finite {--} although often small {--}
upward renormalization of $c_{44}$ even in infinitely thick samples. 
This renormalization is absent in calculations based on the local
boson mapping \cite{TN}. In fact,
in the work of TN
an important role is played by
the invariance of the flux-line interaction under an affine
transformation or uniform tilt  (corresponding to Galilean invariance
of a pure boson system). L.D. Landau \cite{Landau} has shown that the Galilean invariance
implies that the superfluid density at the ground state ($T=0$) of a
superfluid equals the total density. The affine transformation invariance is not present in the more general 
intervortex free energy that allows for 
pairwise interaction among vortex segments at different heights, $z$.
This nonlocality 
breaks the ``Galilean invariance'' 
and yields a  tilt-tilt interaction which
penalizes any misalignment of the flux-lines, therefore favoring
disentanglement.

\vskip .2in
M.C.M. has benefitted from conversations with David Nelson and Steven Teitel.
This work was supported by the National Science Foundation at Syracuse
through Grants DMR-9730678 and DMR-9805818.

\end{multicols}

\section{Appendix A - Derivation of nonlocal hydrodynamics from the 
partition function of $2d$ charged bosons}

In this appendix we show that the nonlocal, non-Gaussian
hydrodynamic free energy given in Eq. (\ref{nonlinhydF})
can be derived by formal manipulations of 
the partition function of
a two-dimensional charged boson fluid. 
Feigel'man and collaborators \cite{feigel} have
shown that the partition function of an array of flux-lines described in the London approximation by the 
Ginzburg-Landau free
energy of Eq. (\ref{gibbs0}) can be mapped onto that of a two-dimensional system of bosons
interacting  via a massive vector potential. The nonlocality of the intervortex interaction
is incorporated via a gauge field that mediates a retarded interaction among the bosons.
The coherent-state formulation of the boson problem yields the
imaginary-time action:
\begin{eqnarray}
\label{Feig}
{\cal S}_c[{\psi},\psi^*,{\bf a},{\bf A}] &=&
   {\int}_{0}^{{\beta}{\hbar}}d{\tau}{\int}d{\bf r}_{\perp}\big\{\;{\psi}^*[\;\hbar\partial_{\tau}+ia_{0}-{\frac{1}{2m}}(\hbar{\bbox
\nabla}_{\perp}+i{\bf
a}_{\perp})^2-\mu\;]{\psi}+V_{sr}({\psi}{\psi}^*)+{\frac{p^2}{2g^2}}({\bbox
\nabla}_{\perp}\times{\bf a}_{\perp})^2+ \nonumber\\
& & {1\over{2g^2}}\big[\hat{\bf z}\times(\partial_\tau{\bf a}_\perp-\bbox{\nabla}a_0)\big]^2
  +{\frac{i}{2\sqrt{\pi}{\tilde\lambda_\perp}{g}}}({\bbox
\nabla}\times{\bf a})\cdot{\bf A}+{\frac{1}{8{\pi}}}({\bbox
\nabla}\times{\bf A})^2\; \big\}\;.
\end{eqnarray}
The correspondence between vortex and boson variables is summarized in Eq. (\ref{bosonmap}).
The coupling constant $g$ corresponds to the strength of the vortex interaction,
according to $g^2\leftrightarrow \phi_0^2/(4\pi\tilde\lambda_\perp^2)$
and $p$ is the anisotropy parameter that here allows for different 
scalar an transverse interaction among the bosons.
${\bf A}$ is the vector potential of the real magnetic field $({\bbox
\nabla}\times{\bf A}={\bf B})$,  and ${\bf a}=(a_{0},{\bf a}_{\perp})$ is a  gauge
field that mediates the non-instantaneous interaction among the
bosons. The boson chemical potential $\mu$ has to be determined so that
the equilibrium boson density $n_B$ equals the vortex density,
$n_B=n_0=B/{\phi}_{0}$. Finally, 
$V_{sr}$ is a short range repulsion
(on scale ${\xi}$) between the bosons. This action is based on the gauge
${\bbox \nabla}\cdot{\bf A}=0$ and ${\bbox \nabla}_{\perp}\cdot{\bf
a}_{\perp}=0$. The choice of ${\bbox \nabla}_{\perp}\cdot{\bf
a}_{\perp}=0$ instead of ${\bbox \nabla}\cdot{\bf a}=0$ reflects the
assumption of  nonrelativistic velocities for the bosons, corresponding to small tilt
of the flux lines away from the $z$ direction \cite{noterel}.
By rewriting the boson fields in 
terms of an amplitude and
a phase, as defined in Eq. (\ref{amplphase}), we obtain
\begin{eqnarray}
\label{Feigntheta}
{\cal S}_c[\hat{n},\theta,{\bf a},{\bf A}]&=&\int_{0}^{\beta{\hbar}}d{\tau}\int
d{\bf r}_{\perp}\big\{\;i\hbar\hat{n} \partial_{\tau}\theta + i \hat{n} a_{0} +
{\frac{\hbar^2}{8m}}{\frac{({\bbox \nabla}_{\perp} \hat{n})^2}{\hat{n}}} +
V_{sr}(\hat{n}) + 
{\hat{n}\over{2m}}{\bf a}_{\perp}^2 + {\hbar\over{m}}\hat{n}({\bbox
\nabla}\theta)\cdot{\bf a}_{\perp}-\mu \hat{n} +\nonumber\\
& & {{\hbar}^2\over{2m}}n({\bbox \nabla}_{\perp}\theta)^2 + 
{\frac{p^2}{2g^2}}({\bbox
\nabla}_{\perp}\times{\bf a}_{\perp})^2 
+{1\over{2g^2}}\big[\hat{\bf z}\times(\partial_\tau{\bf a}_\perp-\bbox{\nabla}a_0)\big]^2+\nonumber\\
& &{\frac{i}{2\sqrt{\pi}{\tilde\lambda_\perp}{g}}}({\bbox \nabla}\times{\bf
a})\cdot{\bf A}+{\frac{1}{8\pi}}({\bbox \nabla}\times{\bf A})^2\; \big\}\;.
\end{eqnarray}
The assumption of small fluctuations allows us to extend the range of
$\theta$ from $[-\pi,\pi]$ to $[-\infty,+\infty]$. As described in section V,
we now eliminate the phase $\theta$
in favor of a vector field $\hat{\bf P}$ via a Hubbard-Stratonovich transformation,
to obtain
\begin{eqnarray}
\label{PnFeig}
\tilde{\cal S}'_c[\hat{n},\hat{\bf P},{\bf a},{\bf A}]&=&\int_{0}^{\beta\hbar} d{\tau}\int
d{\bf r}_{\perp} \big\{\;{\frac{\hbar^2}{2m}}\hat{n}\hat{\bf P}^2 +
{\frac{\hat{n}}{2m}}{\bf a}_{\perp}^2+
i \hat{n} a_{0} +
{\frac{\hbar^2}{2m}}{\frac{({\bbox \nabla}_{\perp}\hat{n})^2}{\hat{n}}} -{\mu}\hat{n}
+V_{sr}(\hat{n}) +\nonumber\\
& &  {\frac{p^2}{2g^2}}({\bbox
\nabla}_{\perp}\times{\bf a}_{\perp})^2 
+{1\over{2g^2}} \big[\hat{\bf z}\times(\partial_\tau{\bf a}_\perp-\bbox{\nabla}a_0)\big]^2
+{\frac{i}{2\sqrt{\pi}{\tilde\lambda_\perp}{g}}}({\bbox \nabla}\times{\bf a})\cdot{\bf
A}+\nonumber\\
& &{1\over{8{\pi}}}({\bbox \nabla}\times{\bf A})^2+
{\frac{n_{0}{\hbar^2}}{m}}\ln({\hat{n}\over n_{0}})\;\big\},
\end{eqnarray}
with the constraint
\begin{eqnarray}
\label{Feigconstr}
\partial_{\tau}\hat{n} + {\bbox \nabla}_{\perp}\cdot{\hat{n}\over m}(\hbar\hat{\bf P}+i
{\bf a}_{\perp})=0\;.
\end{eqnarray}
The last term in the action in Eq. (\ref{PnFeig}), logarithmic in the density,
is the Jacobian of the transformation.
We then make a change of variables,
\begin{equation}
\label{tiltFeig}
\hat{\bf t}={\hat{n}\over m}(\hbar\hat{\bf P}+i{\bf a}_{\perp})\;,
\end{equation}
and obtain
\begin{eqnarray}
\label{Feigeffprime}
{\cal S}'_c[\hat{n},\hat{\bf t},{\bf a},{\bf A}]&=& \int_{0}^{\beta\hbar} d{\tau}\int 
d{\bf r}_{\perp}\big\{ {\frac{m\hat{\bf t}^2}{2\hat{n}}} - i {\bf a}_{\perp}\cdot\hat{\bf
t} + i \hat{n} a_{0} - \mu \hat{n} + V_{sr}(\hat{n}) +{\frac{\hbar^2}{8m}}{\frac{({\bbox
\nabla}_{\perp}\hat{n})^2}{\hat{n}}} +\nonumber\\ 
& & {\frac{p^2}{2g^2}}({\bbox
\nabla}_{\perp}\times{\bf a}_{\perp})^2 
+{1\over{2g^2}}\big[\hat{\bf z}\times(\partial_\tau{\bf a}_\perp-\bbox{\nabla}a_0)\big]^2
+{\frac{i}{2\sqrt{\pi}{\tilde\lambda_\perp}{g}}}({\bbox \nabla}\times{\bf a})\cdot{\bf
A}+
{1\over{8{\pi}}}({\bbox \nabla}\times{\bf A})^2\;\big\},
\end{eqnarray}
with the constraint
\begin{eqnarray}
\label{cons}
\partial_{\tau} \hat{n} + {\bbox \nabla}_{\perp}\cdot\hat{\bf t} = 0\;.
\end{eqnarray}
The Jacobian of this transformation cancels that of the previous one. 

Finally, we define an effective action ${\cal S}_c^{\rm eff}$ for the bosons by integrating out both
the vector potential  ${\bf A}({\bf r})$ and the gauge field
${\bf a}({\bf r})$, 
\begin{equation}
\int' {\cal D}\hat{n}{\cal D}\hat{\bf t}{\cal D}{\bf A}{\cal D}{\bf a}
~e^{-{\cal S}'_c[\hat{n},\hat{\bf t},{\bf a},{\bf A}]}
\delta(\partial_{\tau} \hat{n} + {\bbox \nabla}_{\perp}\cdot\hat{\bf t})=
\int {\cal D}\hat{n}{\cal D}\hat{\bf t}
e^{-{\cal S}^{\rm eff}_c[\hat{n},\hat{\bf t}]}
\delta(\partial_{\tau} \hat{n} + {\bbox \nabla}_{\perp}\cdot\hat{\bf t}).
\end{equation}
The prime over the integral sign on the left hand side of the equation indicates
that the integration over ${\bf A}$ and ${\bf a}$ has to be performed by taking into
account the constraints imposed by our choice of gauge.
The vector potential and gauge field are most easily integrated out by rewriting
the field part of the action (\ref{Feigeffprime}) in Fourier space,
with the result,
\begin{eqnarray}
\label{Feighyd}
{\cal S}^{\rm eff}_c[\hat{n},\hat{\bf t}]= & &
  \int_{0}^{\beta\hbar}d\tau\int d{\bf r}_{\perp} \big\{\;{m\hat{\bf
t}^2\over{2\hat{n}}} - \mu \hat{n} + V_{sr}(\hat{n}) + {{\hbar}^2\over{8m}}{({\bbox
\nabla}_{\perp}\hat{n})^2\over{\hat{n}}}\;\big\} \nonumber\\
 & & + {1\over{2\Omega}}\sum_{\bf
q}\big\{\;{{g^2{\tilde\lambda_\perp}^2}\over{1+q_z^2{\tilde\lambda_\perp}^2
   +q_{\perp}^2p^2{\tilde\lambda_\perp}^2}}|\hat{\bf t}_{T}({\bf q})|^2 +
  {q^2\over q_{\perp}^2}{g^2\tilde\lambda_\perp^2\over{1+q^2\tilde\lambda_\perp^2}}|\hat{n}({\bf q})|^2\;\big\},
\end{eqnarray}
where $\hat{\bf t}_T({\bf q})=\hat{\bf q}_\perp\times\hat{\bf t}({\bf q})$. 
By making use of the continuity constraint given in Eq. (\ref{cons}),
we can write
\begin{eqnarray}
{g^2\tilde\lambda_\perp^2\over1+q_z^2\tilde\lambda_\perp^2+q_\perp^2p^2\tilde\lambda_\perp^2}
|\hat{\bf t}_{T}({\bf q})|^2
+ {q^2\over q^2_\perp}{{g^2{\tilde\lambda_\perp}^2}\over{1+q^2\tilde\lambda_\perp^2}}|\hat{n}({\bf q})|^2 
=& &{{g^2{\tilde\lambda_\perp}^2}\over{1+q_z^2+q_{\perp}^2p^2{\tilde\lambda_\perp}^2}}|\hat{\bf t}({\bf q})|^2
+\nonumber\\
& &{{g^2{\lambda}^2}(1+q^2p^2{\tilde\lambda_\perp}^2)\over{(1+q^2{\tilde\lambda_\perp}^2)
(1+q_z^2{\tilde\lambda_\perp}^2+q_{\perp}^2p^2{\tilde\lambda_\perp}^2)}}
|\hat{n}({\bf q})|^2.
\end{eqnarray}
Finally, if we replace the short range repulsion $V_{sr}(\hat{n})$ by a 
short-wavelength cutoff and
identify the boson density $\hat{n}$ and momentum field $\hat{\bf t}$
with the corresponding hydrodynamic quantities for the vortices, 
we see that Eq. (\ref{Feighyd}) yields precisely the nonlocal non-Gaussian hydrodynamic free 
energy discussed in section VI.

\section{Appendix B - Perturbative corrections to the tilt modulus from nonlinear hydrodynamics}

The wave-vector dependent tilt modulus is defined in terms of the transverse
part of the tilt-tilt correlator as in
Eq. (\ref{longtilt}). In the hydrodynamic approximation, the tilt-tilt correlator can be written as
\begin{equation}
\label{ttcor}
T_{ij}({\bf r},{\bf r}')=\frac{\int{\cal D}\hat{n}({\bf r}){\cal D}\hat{\bf t}({\bf
r})\hat{t}_{i}({\bf r})\hat{t}_{j}({\bf r}')e^{-F/k_BT} \delta(\partial_{z}\hat{n}+{\bbox
\nabla}_{\perp}\cdot\hat{\bf t})}{\int{\cal D}\hat{n}(\bf r){\cal D}\hat{\bf t}({\bf
r}) e^{-F/k_BT} \delta(\partial_{z}\hat{n}+{\bbox
\nabla}_{\perp}\cdot\hat{\bf t})}\;,
\end{equation}
Where $F$ is the hydrodynamic free energy given in Eq. (\ref{nonlinhydF}). 
The free energy can be written as the sum of a Gaussian part, $F_{G}$, and non-Gaussian 
corrections, $\delta F$, as in Eq. (\ref{splitF}). We want to calculate up
to lowest-order in the small parameter, $u^2$, nonlinear corrections to the
tilt autocorrelator. 
By keeping only terms up to fourth order
in the fluctuations of the hydrodynamic fields, the non-Gaussian part of the free energy is given by,
\begin{eqnarray}
\delta  F \approx  - \frac{\tilde{\epsilon}_1}{2 n_0^2
\Omega^2 }\sum_{{\bf q}_1, {\bf q}_2}  \hat{t}_i({\bf q}_1) \hat{t}_i({\bf q}_2) \delta\hat{n}
(-{\bf q}_1 - {\bf q}_2)  +
\frac{\tilde{\epsilon}_1}{2 n_0^3
\Omega^3} \sum_{{\bf q}_1, {\bf q}_2, {\bf q}_3}  \hat{t}_i({\bf q}_1) \hat{t}_i ({\bf
q}_2)\delta\hat{n} ({\bf q}_3) \delta\hat{n} (-{\bf q}_1-{\bf q}_2-{\bf q}_3) \;. 
\end{eqnarray}
The tilt-tilt correlator is then evaluated in Fourier space perturbatively
in the non-Gaussian part of the free energy, with the result,
\begin{equation}
\label{ttexpan}
T_{ij}({\bf q},{\bf q}')=\Omega\delta_{{\bf
q}+{\bf q}',{\bf 0}}  T_{ij}^{0}({\bf q})-{1\over k_BT}\langle \hat{t}_{i}({\bf q})\hat{t}_{j}({\bf 
q}')\delta F \rangle_{G}^c
+{\frac{1}{2(k_BT)^2}}\langle  \hat{t}_{i}({\bf q})\hat{t}_{j}({\bf q}')(\delta F)^2 
\rangle_{G}^c\;,
\end{equation}
where $\langle ...\rangle_G^c$ denotes a cumulant average over the 
Gaussian ensemble with weight $\sim\exp(-F_G/k_BT)$. 
The first term on the right hand side of Eq. (\ref{ttexpan}) is the Gaussian
result given in Eqs. (\ref{elcorI}-\ref{longittilt}). 

Using Wick's theorem, the corrections arising from the non-Gaussian part of the free energy are easily
expressed in terms of the correlations in the Gaussian ensemble given in Eq. 
(\ref{elcorII}-\ref{longittilt}), with the result
\begin{equation}
P_{ij}^{T}({\bf q}_{\perp}) \langle \hat{t}_{i}({\bf q})\hat{t}_{j}({\bf q}')\delta F 
\rangle_G^c=\Omega\delta_{{\bf
q}+{\bf q}',{\bf 0}}[T_{T}^{0}({\bf q})]^2{\frac{\tilde{\epsilon}_1}{n_0^3}}{\frac{1}{\Omega}}\sum_{{\bf
q}_{1}}\langle|\delta \hat{n}({\bf q}_{1})|^2\rangle_{G}
\end{equation}
and
\begin{eqnarray}
P_{ij}^{T}({\bf q}_{\perp}) \langle \hat{t}_{i}({\bf q})\hat{t}_{j}({\bf q}')(\delta
F)^2 \rangle_G^c= & & 2\Omega\delta_{{\bf
q}+{\bf q}',{\bf 0}}[T_{T}^{0}({\bf q})]^2P_{ij}^{T}({\bf q}_\perp){\frac{\tilde{\epsilon}_1^2}{n_0^4}}
{\frac{1}{\Omega}}\sum_{{\bf q}_{1}} \Big\{\langle \hat{t}^H_{i}({\bf q}_1) \hat{t}^H_{j}(-{\bf
q}_1)\rangle_G\langle|\delta\hat{n}^H({\bf q}-{\bf q}_1)|^2 \rangle_G +\nonumber\\
& &\langle \hat{t}^H_{i}({\bf q}_1)\delta\hat{n}^H(-{\bf q}_1)\rangle_G\langle \hat{t}^H_{j}({\bf
q}_1-{\bf q}) \delta \hat{n}^H({\bf
q}-{\bf q}_1)\rangle_G\Big \}.
\end{eqnarray}
By substituting the expressions for the Gaussian correlators given 
in Eqs. (\ref{elcorII}-\ref{longittilt}), we obtain the following expression for the
transverse part of the tilt autocorrelator to lowest order in the non-Gaussian terms,
\begin{eqnarray}
T_{T}({\bf q})=& & \frac{n_{0}^2 k_{B}T}{c_{44}^{0}({\bf
q})}\\\nonumber
& &-{\frac{n_{0}^3\tilde{\epsilon}_{1}(k_{B}T)^2}{[c_{44}^{0}({\bf 
q})]^2}}{1\over LA}\sum_{{\bf q}'_\perp, q'_z}\bigg\{
     {{q'}_\perp^2\over c_{44}^0({\bf q'})}
     {1\over {q'}_{z}^2+[\xi_z({\bf q'})]^{-2}} - 
    {n_0\tilde{\epsilon}_1({\bf q}_\perp -{\bf q}'_\perp)^2
           \over c_{44}^0({\bf q'})
                 c_{44}^0({\bf q}-{\bf q}')}
     {1\over (q_z-q'_z)^2+[\xi_z({\bf q}-{\bf q'})]^{-2}}\bigg\} \\ \nonumber
& &  -{n_0^4{\tilde{\epsilon}_1}^2(k_BT)^2\over{[c_{44}^{0}({\bf q})]^2} 
}{1\over{LA}}\sum_{{\bf q}'_\perp, q'_z}
     {({\bf\hat{q}}_\perp\cdot{\bf\hat{q}'}_\perp)^2({\bf q}_\perp-{\bf 
q'}_\perp)^2
     [\xi_z({\bf q'})]^{-2} 
        -\big[1-({\bf\hat{q}}_\perp\cdot{\bf\hat{q}'}_\perp)^2\big]{q'}^2_\perp
              {q'}_z({q'}_z-q_z)\over c_{44}^0({\bf q'})
                    c_{44}^0({\bf q}-{\bf q'})
     \big[{ q'}_{z}^2+[\xi_z({\bf q'})]^{-2}\big]
      \big[ (q_z-q'_z)^2+[\xi_z({\bf q}-{\bf q'})]^{-2}\big]}\;.
\end{eqnarray}
\begin{multicols}{2}

\end{multicols}


\begin{references}
\bibitem{foota}
Another possibility is that the longitudinal gauge symmetry is recovered 
first, while the translational symmetry remains broken, yielding a
``supersolid'' phase.
\bibitem{factoroftwo}
There is a discrepancy of a factor of two between our result for the normal-fluid
fraction in the limit $v^*=0$ and the result of Ref. \onlinecite{feigel}.
We believe that this is due to a typo in Ref. \onlinecite{feigel}.
\bibitem{noterel}
In this nonrelativistic approximation we can use the $2D$ Coulomb gauge
rather than the more general Lorentz gauge.
\end{references}
\end{document}